\documentclass[fleqn,usenatbib]{mnras}

\usepackage{newtxtext,newtxmath}

\usepackage[T1]{fontenc}

\DeclareRobustCommand{\VAN}[3]{#2}
\let\VANthebibliography\thebibliography
\def\thebibliography{\DeclareRobustCommand{\VAN}[3]{##3}\VANthebibliography}

\newcommand{\vb}{v_\mathrm{broad}}
\newcommand{\teff}{T_\mathrm{eff}}
\newcommand{\grvs}{G_\mathrm{RVS}}
\newcommand{\bp}{G_\mathrm{BP}}
\newcommand{\rp}{G_\mathrm{RP}}
\newcommand{\vsini}{v\sin i}
\newcommand{\dc}{\Delta\,\mathrm{Color}}
\newcommand{\logrhk}{\log\, R^\prime_\mathrm{HK}}
\newcommand{\wcrit}{\omega_\mathrm{crit}}

\usepackage{graphicx}	
\usepackage{amsmath}	
\usepackage{adjustbox}
\usepackage[table]{xcolor}

\title[Stellar rotation in Galactic open clusters]{Survey of extended Main Sequence Turn-offs in Galactic Open Clusters: Stellar rotations from Gaia RVS spectra}

\author[G. Cordoni et al.]{
G. Cordoni$^{1,2 }$,\thanks{E-mail: giacomo.cordoni@anu.edu.au}
L. Casagrande$^{1, 2}$,
J. Yu$^{1, 2, 3}$,
A. P. Milone$^{4, 5}$, 
A. F. Marino$^{5, 6}$,
F. D'Antona$^{7}$, 
\newauthor  
F. Dell'Agli$^{7}$,
S. Buder$^{1, 2}$,
and M. Tailo$^{8}$ 
\\
$^{1}$ Research School of Astronomy and Astrophysics, The Australian National University, Canberra, ACT 2611, Australia \\
$^{2}$ARC Centre of Excellence for All Sky Astrophysics in 3 Dimensions (ASTRO 3D), Australia\\
$^{3}$ School of Computing, Australian National University, Acton, ACT 2601, Australia \\
$^{4}$ Dipartimento di Fisica e Astronomia''Galileo Galilei`` - Univ. di Padova, Vicolo dell'Osservatorio 3, Padova, IT-35122 \\
$^{5}$ Istituto Nazionale di Astrofisica - Osservatorio Astronomico di Padova, Vicolo dell'Osservatorio 5, Padova, IT-35122 \\
$^{6}$ Istituto Nazionale di Astrofisica - Osservatorio Astrofisico di Arcetri, Largo Enrico Fermi, 5, Firenze, IT-50125 \\
$^{7}$ Istituto Nazionale di Astrofisica, Osservatorio Astronomico di Roma, Via Frascati 33, 00077 Monte Porzio Catone, Italy \\
$^{8}$ Dipartimento di Fisica e Astronomia Augusto Righi, Univ. degli Studi di Bologna, Via Gobetti 93/2, 40129 Bologna, Italy
}

\date{Accepted 2024 June 20. Received 2024 June 19; in original form 2024 February 26}

\pubyear{2024}

\begin{document}
\label{firstpage}
\pagerange{\pageref{firstpage}--\pageref{lastpage}}
\maketitle

\begin{abstract}
The origin of extended main-sequence turn-offs (eMSTO) in star clusters younger than 2 Gyr still challenges our current understanding of stellar evolution. Exploiting data from Gaia Data Release 3 (DR3), we investigate eMSTOs in a large sample of 32 Galactic open clusters younger than 2.4\,Gyr.
We first validate Gaia rotational velocities from Radial Velocity Spectrometer (RVS) spectra by comparing them with literature values and assessing their correlation with magnetic activity measurements from LAMOST spectra. We detect a general positive correlation between turn-off color and projected stellar rotation, with slow-rotating stars predominantly found on the bluer side of the turn-off. Comparing our observations with theoretical models, we find that the eMSTO morphology is well-reproduced by a single population formed with a high rotation rate, and observed with rotation axis inclination ranging between 0$^\circ$ (pole-on) and 90$^\circ$ (edge-on). This contrasts with observations of Magellanic Clouds clusters, where a population of non-rotating stars appears to be ubiquitous in clusters younger than 700\,Myr. However, we note that our interpretation, while successfully explaining the overall eMSTO morphology, cannot fully explain the observed projected rotational velocities. Additionally, two young clusters, NGC\,3532 and NGC\,2287, exhibit moderate evidence of a split main sequence in color and rotation, suggesting a possible small spread in the initial rotation rate.
Finally, we advise caution in determining the ages of young clusters from non-rotating isochrones, as neglecting the effects of stellar rotation can impact the isochrone dating by up to factors of 5-20\%.

\end{abstract}




\section{Introduction}
\label{sec:intro}
High-precision data from the Hubble Space Telescope (HST) have revealed intriguing features in the color-magnitude diagrams (CMDs) of star clusters younger than approximately 2\,Gyr in the Large and Small Magellanic Clouds (LMC and SMC). Nearly all young clusters exhibit extended main-sequence turnoffs (eMSTOs), and those younger than about 700 million years display both eMSTOs and split main sequences \citep[MSs,][and series]{bertelli2003, mackey2007, milone2009, milone2023a}. The comparison between observed CMDs and theoretical models \citep[e.g the Geneva stellar models,][]{georgy2013} suggests that the split MSs are consistent with two stellar populations having different rotation rates. Specifically, stars populating the red MS are interpreted as stars with rotation close to the breakout velocity, while stars on blue MS are consistent with being slow or non-rotators. On the other hand, theoretical isochrones predict that in clusters younger than 0.5-1 Gyr, slow or non rotating stars would evolve populating the faint and red end of the turn-off, while fast rotators would move toward brighter magnitudes and bluer colors \citep[see e.g.][for a detailed description]{niederhofer2015, dantona2017, milone2018}. A visual representation of the phenomenon is shown in Fig.~\ref{fig:iso parsec}, where Parsec isochrones \citep{marigo2017, girardi2019, nguyen2022} computed for different ages, initial rotation rates and inclinations are displayed. A detailed description is provided in App.~\ref{app:parsec}.

Furthermore, the presence of Be-stars, i.e. very rapidly rotating stars surrounded by a decretion disc and characterized by a strong $H_{\rm \alpha}$ emission and possibly dust, further increases the complexity of the eMSTO phenomenon in young Magellanic Clouds clusters \citep[we refer to ][for a detailed description of Be stars and the debated connection with interstellar dust]{milone2018, milone2023b, milone2023a, dantona2023, kamann2023, martocchia2023}.

In recent years, the precise photometric and astrometric data from the Gaia mission have revealed that eMSTOs and split main-sequences are not unique to young and massive Magellanic Clouds clusters but are also widespread among young Galactic Open clusters \citep[OCs,][]{marino2018b, bastian2018, cordoni2018, li2024, bu2024}. This ubiquity across clusters younger than 2 Gyr suggest a likely link with stellar and cluster evolution, thus underscoring the significance of the phenomenon. However, the precise nature and mechanisms behind these observed features remain unknown. Proposed explanations include: \textit{(i)} the idea that all stars form as fast rotators and are subsequently braked by magnetic and/or tidal forces \citep{dantona2015}; \textit{(ii)} the intrinsic bimodal rotation rates distribution, linked to star formation and dependent on pre-Main Sequence stars' disc lifetime \citep{bastian2020}; \textit{(iii)} the hypothesis that stellar mergers and different evolutionary pathways can account for observed eMSTO and split-MS \citep{wang2022}; \textit{(iv)} the presence of circumstellar dust and the consequent light absorption, which can potentially contribute to shaping young clusters' CMDs \citep{dantona2023}.

Since the first discovery of eMSTOs and split MSs, many developments have been made in the field. Specifically, direct evidence confirming the presence of different rotators in the red and blue MSs have been provided using spectroscopic information collected through the Very Large Telescope and Magellan-Clay telescope\citep[VLT, see e.g.][respectively]{dupree2017, marino2018a} and MUSE \citep{kamann2020, kamann2023}. 
Nonetheless, quantitative comparisons between theoretical models and observed CMDs yield inconclusive proof on whether rotation alone is the driving mechanism or if other mechanisms contribute, like prolonged star formation episodes are required to reproduce the observations \citep{gossage2019, lipatov2022}. Specifically, the color spread in the turn-off can be reproduced by a population of rapidly rotating stars, but either a younger or braked population, may be required to reproduce the bluest end of the eMSTO in young clusters \citep{dantona2015, dantona2017, milone2017}. Conversely, \citet{cordoni2022} investigated the young LMC cluster NGC\,1818, excluding age differences larger than 5 Myr.

Furthermore, \citet[][]{milone2023b, milone2023a, martocchia2023, dantona2023} have identified and investigated a new and unique population of stars, dubbed as UV-dim stars, whose photometry is strongly absorbed in the UV regime but minimally affected in the F814W HST near-infrared band. The different works concluded that UV-dim stars are present in young ($\sim 100$ Myr) and intermediate age ($\sim 1.5 Gyr$) cluster alike \cite[][]{milone2023b, milone2023a}, however, the exact nature of UV-dim stars remains enigmatic \citep[see e.g.,][]{milone2023b, kamann2023, martocchia2023, dantona2023}.

The exact nature of the eMSTO phenomenon remains elusive, thus hindering our understanding of the formation and evolution of seemingly ``simple'' young star clusters. While considerable research has focused on massive young clusters in the Magellanic Clouds, only a limited number of studies have targeted Galactic Open clusters. This study aims to bridge this gap by conducting the first homogeneous and systematic analysis of eMSTOs in Galactic young star clusters exploiting Gaia Data Release 3 \citep[DR3,][]{gaiadr3}, 
and LAMOST Data Release 9 \citep[DR9 v2.0,][]{cui2012}. By combining Gaia DR3 photometry and the recently published line broadening velocities \citep[$\vb$, ][]{fremat2023}, which serve as proxies of projected stellar rotation, with a stellar magnetic activity index determined from LAMOST DR9, we will study the morphology of the eMSTO and contribute fundamental insights toward understanding this phenomenon.

The paper is structured as follows: in Sec.~\ref{sec:data} we will discuss the initial dataset and the data validation procedure, Sec.~\ref{sec:toff} will discuss the photometric properties of the eMSTO in the whole of clusters while the results will be presented in Sec.~\ref{sec:results}. Finally, discussion and summary will be presented in Sec.~\ref{sec:discussion} and \ref{sec:conclusions} respectively.

\section{Dataset and $\vb$ validation}
\label{sec:data}

In recent years, the exquisite photometry and astrometry provided by the Gaia mission, and in particular Gaia DR3 \citep{gaiadr3} revolutionized our understanding of the Milky Way, shedding light on its formation and accretion history, as well as on its star clusters population. Specifically, many works focused on the detection and characterization of open clusters by means of different techniques \citep[see e.g.][]{cantat2018a, cantat2020, cantat2022, bossini2019, tarricq2021, dias2021, hunt2024}. In this work, we will exploit the membership and reddening information provided in previous works \citep{cantat2020, cordoni2023} to conduct a homogeneous investigation of the eMSTOs phenomenon and the role of stellar rotation in Galactic OCs. 
For the sake of our analysis, we selected OCs younger than 2.5 Gyr, within 4 kpc and with more than 200 cluster members. An additional selection has been made on the basis of the overall CMD quality. The final sample includes 32 OCs spanning from 40 Myr to 2.5 Gyr, 17 of which have Gaia DR3 $\vb$ available for a sufficient number of turn-off stars. The list of the analyzed clusters, as well as some relevant properties are included in Tab.~\ref{tab:tab1}.

\subsection{Cluster members and reddening correction}
As discussed in \citet{cordoni2023}, cluster members have been selected on the basis of their location in the plane of the sky, proper motions, parallax and position in the CMD. Moreover, the photometry of cluster members has been corrected for differential reddening (DR), i.e. reddening variations across the field of view, following the recipe described in \citet{milone2012} and already adopted in \citet{cordoni2023}. 
In a nutshell, we first calculated the cluster's fiducial line in the CMD by using a sample of bonafide reference stars, composed of bright MS stars. We then used the absorption coefficients for Gaia passbands determined in \citet{casagrande2018} to calculate the color and magnitude displacement of each reference star with respect to the fiducial line, along the reddening direction.
To compute the differential reddening associated with each star, we selected the 25 spatially closest reference stars and calculated the median distance along the reddening line. Additionally, for each star we also determined the resolution of the reddening correction as the maximum distance of the 25 neighboring stars used in the star's reddening  calculation. We refer to \citet{milone2012} and \citet{legnardi2023} for a more detailed description of the procedure. 

\subsection{Line broadening parameter validation}
Recently, Gaia DR3 publicly released line broadening velocities (hereafter referred to as $\vb$) for 3 524 677 stars with $\teff \in (3500, 14500)$ K and $\grvs$ magnitude brighter than 12. These broadening velocities are determined using the the Radial Velocity Spectrometer (RVS, $\lambda/\Delta\,\lambda \sim 11500$) which measures stellar radial velocities using Ca Triplet lines in the range 846-870\,nm. We refer to \citet{fremat2023} for a detailed description of Gaia DR3 $\vb$ velocities. The shape of the lines measured by the RVS not only depends on stellar parameters, but also on additional physical (e.g., rotation, macro-turbulence, pulsations, binarity, etc ... ) as well as instrumental effects \citep[e.g. line spread function, see][]{sartoretti2023}. However, the most dominant physical mechanism is stellar axial rotation which produce significant line broadening due to Doppler effect. The latter depends on the intrinsic stellar rotational velocity $v$, projected by the star inclination angle $i$, i.e. $\vsini$. Specifically, the inclination $i$ range from $i=0^\circ$ (star observed pole-on, no net rotation) and $i=90^\circ$ (star observed edge-on, maximum rotation). As the Gaia pipeline did not quantify the influence of different mechanism on the line broadening, but rather matched the line broadening with a solely rotation induced broadening, the inferred quantity is referred to as $\vb$ and not $\vsini$. Nonetheless, throughout the paper we will consider $\vb$ as a reliable proxy of projected stellar rotation for the reasons discussed in the following paragraphs.  

We followed the recipe described in \citep[][see e.g. their Table 4]{fremat2023} to include in our analysis only stars for which $\vb$ velocities have a high-chance of being consistent with $\vsini$ within $2\/\sigma$. Specifically, \citet{fremat2023} defined a validity domain of the line broadening parameter as a function of $\grvs$ and $\teff$. For each $\grvs$-$\teff$ interval, only $\vb$ in a specific range can be interpreted as reliable proxies of projected stellar rotation. The result of such selection process removes stars across the entire $\vb$ spectrum in clusters older than approximately 0.5\,Gyr, while it removes mostly fast rotators in younger clusters. To test for possible biases introduced by the selection, we repeated the analysis discussed in the next sections without applying the selection criteria, finding consistent results.

As a further check, we compared all Gaia DR3 $\vb$, both selected and excluded, with literature projected rotation velocities. We cross-matched our dataset with GALAH DR3 \citep{buder2021} and \citet{gg2005}, finding consistent values. While the detailed comparison is discussed in App.~\ref{app:galah}, we mention here that $\sim 86 \%$ of stars with available $\vb$ and in common with GALAH DR3 exhibit consistent values, regardless of being selected or excluded by \citet{fremat2023} criteria, while 13\% of the stars display significantly inconsistent projected rotation. However, a detailed analysis of stellar properties and GALAH DR3 normalized spectra \citep[see][for details about spectra acquisition and analysis]{buder2021}, revealed that these stars are likely consistent with being fast rotators, as indicated by Gaia DR3 $\vb$. On the other hand, there is a good agreement between Gaia $\vb$ and \citet{gg2005} across the entire $\vb$ range. We refer to App.~\ref{app:galah} for a more detailed discussion.

\subsection{Stellar activity and stellar rotation} \label{sec:activity}
Standard stellar evolution theory predicts a direct relationship between stellar rotation and chromospheric magnetic activity. Specifically, stellar mass and rotation represent the fundamental parameters shaping the dynamo configuration and, consequently, influencing stellar magnetic activity properties. Conversely, as a star undergoes angular momentum loss and decelerates, its magnetic field weakens, resulting in a decrease in magnetic activity. Observations of various activity tracers, such as stellar X-ray emissions \citep[][]{pizzolato2003}, chromospheric activity and $H_{\rm \alpha}$ emissions \citep{noyes1984, stauffer1997, mamajek2008, douglas2014, fang2018}, consistently revealed clear correlations between rotation, activity, and age.

\begin{figure}
    \centering
    \includegraphics[width=0.50\textwidth, trim={0cm 0cm 0cm 0cm}, clip]{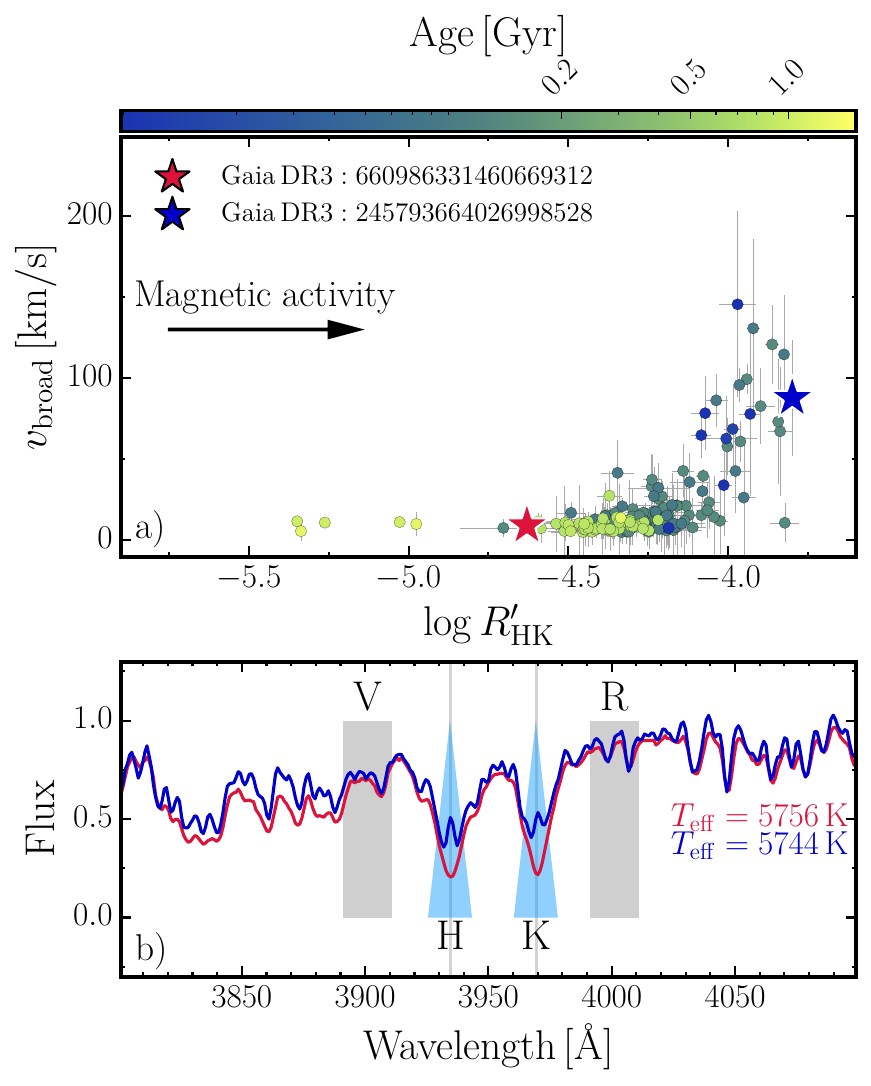}
    \caption{\textit{Panel a)}. Projected stellar rotation vs. magnetic chromospheric activity as indicated by the $\logrhk$ parameter for all stars in common between this study and LAMOST DR9 v2.0 low-resolution data \citep{cui2012}. Stars are color-coded according to the age of the cluster, as indicated in the top colorbar. \textit{Panel b).} LAMOST low-resolution normalized spectra of the red and blue starred symbols in panel b. Spectra are centered around the Ca\,{\rm II} H \& K lines, and their temperature is indicated in the right inset with the same colors. V and R regions are highlighted by the gray rectangles, while the triangular regions used to determine the H \& K emission are represented by azure triangles.}
    \label{fig:rhk} 
\end{figure}

In the present work, we test the reliability of $\vb$ from Gaia by comparing it against chromospheric activity. 
Specifically, to determine stellar activity we adopted the procedure described in \citet{yu2024} to calculate a chromospheric activity index, $\logrhk$, which is based on the Ca\,$\mathrm{II}$ H \& K lines (3968  and 3934 \AA, respectively) from LAMOST DR9 v2.0 low–resolution spectra \citep{cui2012}, and represents the ratio of chromospheric fluxes in Ca\,$\mathrm{II}$ H \& K lines to the bolometric flux. In a nutshell, we integrated emission line fluxes in H and K bandpasses by using 2.18\,\AA\;FWHM triangular windows (azure shaded regions in Fig.~\ref{fig:rhk}b). Such window size ensure that all the emission in Ca\,$\mathrm{II}$ H \& lines are captured in LAMOST spectra with a lower average resolution of 1800. Continuum fluxes for R and V bandpasses (4001 and 3901 \AA) were determined using 20 \AA\;rectangular windows (gray shaded areas in Fig.~\ref{fig:rhk}b).The $S$ index, which is used for calculating $\logrhk$ as in \citet{karoff2016} has been computed as:
\begin{equation}
    S = 1.8 \times 8 \times \frac{2.18\text{\AA}}{20\text{\AA}}\times \frac{H+K}{R+V}.
\end{equation}

The $S$-index estimates have been calibrated by applying the same methodology to stars with known $S$-index and in common with \citet{isaacson2010, isaacson2024, boro2018, gomes2021}.
We then calculated $R'_{\rm HK} = R_{\rm HK} - R_{\rm HK, phot}$, where\footnote{Eq.~\eqref{eq:eq2} is taken from \citet{noyes1984}, Eq.~\eqref{eq:eq3} is derived as in \citet{rutten1984}, Eq.~\eqref{eq:eq4} follows the approach described in \citet{hartmanm1984} \citet{noyes1984}, and Eq.~\eqref{eq:eq5} as in \citet{huang2015}.}
\begin{align}
    & R_{\rm HK} = 1.34 \times 10^{-4}C_{\rm cf} S \label{eq:eq2}\\
    & {\rm log} C_{\rm cf} = 0.25(B-V)^3 - 1.33(B-V)^2 + 0.43(B-V) + 0.24 \label{eq:eq3} \\
    & {\rm log} R_{\rm HK, phot} = -4.898 + 1.918(B-V)^2 - 2.893(B-V)^3  \label{eq:eq4} \\
    & (B-V) = 3.964 \times (T_{\rm eff}/10^4)^2 - 7.709\times (T_{\rm eff}/10^4) + 3.738 \label{eq:eq5}
\end{align}

Finally, we restricted the analysis of chromospheric magnetic activity to F- to M-type stars, with temperatures lower than 6800 K. Indeed, for hotter stars, $\logrhk$ would not provide a reliable proxy for chromospheric activity, and would be dominated by rotation and temperature effects \citep[see e.g.][]{schroder2009, borosaikia2018}. Additionally, the relations used to compute $\logrhk$ (i.e. Eq. \ref{eq:eq2} - \ref{eq:eq5}) are only valid for stars with temperature between 3000 and 7000 K approximately.
Figure~\ref{fig:rhk}a illustrates $\vb$ vs. $\logrhk$ for all stars in common between our catalog and LAMOST DR9 v2.0 low-resolution data, color-coded according to cluster age, as derived in \citet{dias2021}. Gray errorbars represent uncertainties in both $\vb$ and $\logrhk$. The procedure used to compute the S-index is qualitatively represented in Fig.~\ref{fig:rhk}b, where we show the LAMOST DR9 continuum-normalized low-resolution spectra of two stars with low/high magnetic activity, marked by large red/blue starred symbols in Fig.~\ref{fig:rhk}a. The spectral regions used to integrate H \& K emission lines and continuum R \& V fluxes are indicated by the azure and gray shaded regions respectively. Clearly, the spectra of the more active star, i.e. the blue spectra, exhibit a clear peak in the core of both H \& K lines, while no peak is detectable in in the spectra of the less active star.

Figure~\ref{fig:rhk} highlights a clear correlation between magnetic activity (increasing from left to right) and stellar rotation, as faster rotators generally exhibit higher stellar chromospheric activity. These findings align with the results of \citet{xiang2018}, who utilized LAMOST data to explore chromospheric activity in open clusters. Additionally, we note that, as found in \citet{pizzolato2003}, the relation start to saturate for fast rotation rates.
Furthermore, the observed scatter in the relation for high magnetic activities, i.e., $\logrhk \lesssim -4.5$, may be influenced by the inclination of the rotation axis. Indeed, while $\logrhk$ is sensitive to stellar rotation, $\vb$ reflects the projected stellar rotation. Additionally, we note a potential dependency of the relation on cluster age, with younger clusters (darker points) showing a slightly different trend. However, as this is beyond the scope of the present study, we defer further characterization of this phenomenon to future research. 

Overall, the comparison with GALAH DR3 \citep{buder2021} and \citet{gg2005} projected 
velocities, and the relationship between Gaia DR3 $\vb$ and chromospheric stellar activity ($\logrhk$), 
strongly support the conclusion that Gaia $\vb$ provide a reliable estimate of projected stellar rotation. In the following sections, we will therefore consider $\vb$ as proxy projected stellar rotation.

\section{Properties of extended Main-Sequence Turn-Offs}
\label{sec:toff}

In the recent years, different works have demonstrated how the turn-off color spread observed in nearly all clusters younger than $\sim 2$ Gyr in both the Milky Way and Magellanic Clouds can be, partially or completely, attributed to the presence of slow and fast rotators among cluster stars \citep[see e.g.][]{milone2009, bastian2009, goudfrooij2014, niederhofer2015, dantona2015, cordoni2018, gossage2018}. 
In the following, we investigate the morphology of the turn-off color distributions in a sample of 32 clusters. 

\subsection{Turn-Off stars selection}\label{subsec:toff selection}
\begin{figure*}
    \centering
    \includegraphics[width=0.99\textwidth, trim={0cm 0cm 0cm 0cm}, clip]{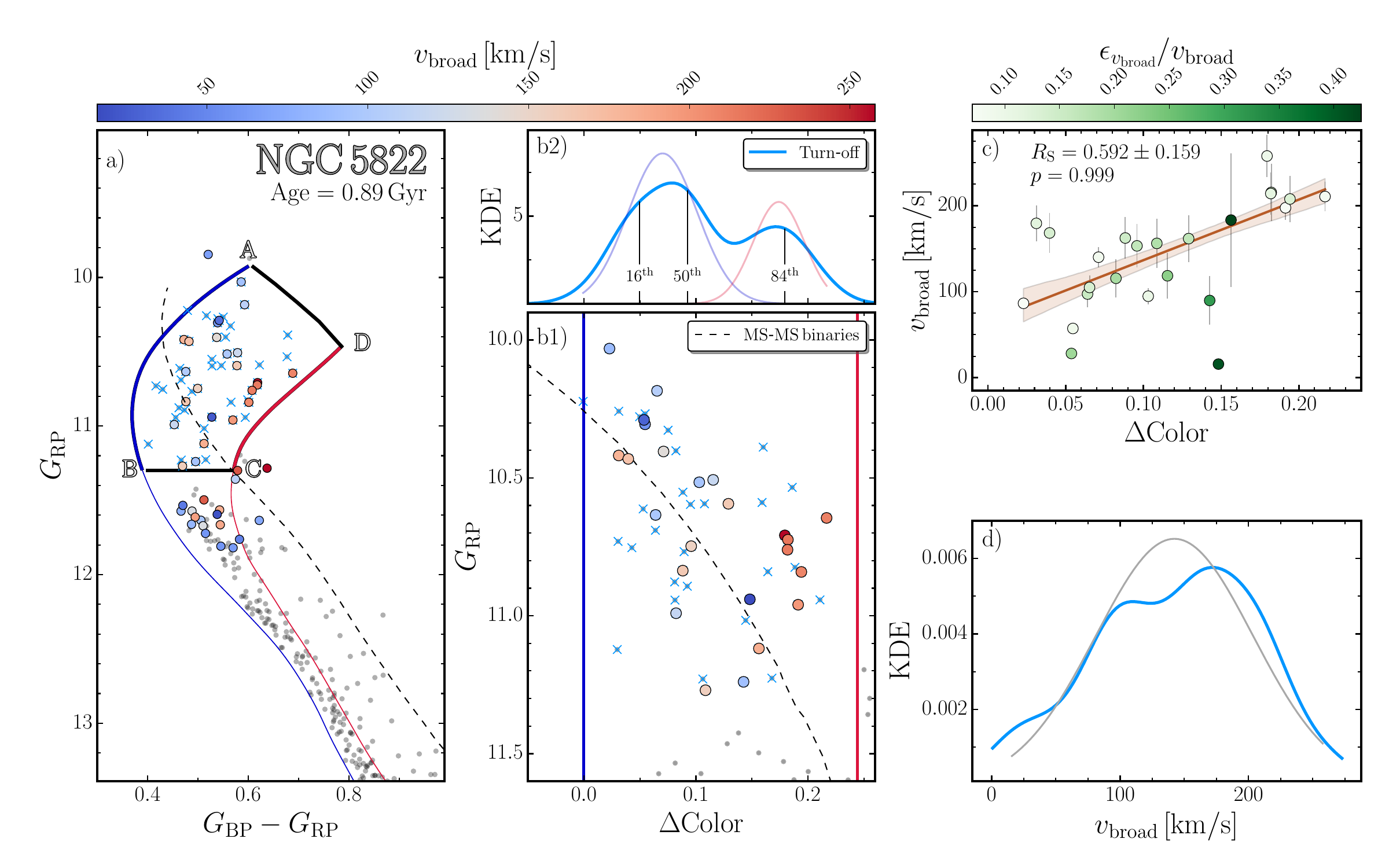}
    \caption{\textit{Panel a).} $\rp$ vs. $\bp - \rp$ CMD of NGC\,5822, where stars have been color-coded according to their $\vb$, if available. Blue and red solid lines indicated the two fiducial lines used to verticalize the color distribution, as discussed in Sec.~\ref{sec:toff}. The turn-off region is defined as the region enclosed by the points ABCD, while the location of MS-MS equal mass binary stars is shown as a dashed black line. Stars selected as turn-off stars are marked with azure crosses. \textit{Panel b1).} Verticalized CMD of turn off stars. Colors and markers are the same as in panel a). \textit{Panel b2).} Distribution of verticalized color ($\dc$) for turn-off stars. The azure solid thick line represents the KDE, while the blue and red shaded lines indicate the best results of the Gaussian Mixture Models fit. The $16^\mathrm{th}, 50^\mathrm{th}$ and $84^\mathrm{th}$ percentiles of the MSTO stars color distribution are indicated by the black vertical lines.  \textit{Panel c).} $\vb$ vs. $\dc$ of turn-off stars with validated $\vb$. Each star has been color coded according to its relative uncertainty, as indicated in the top colorbar. The red line indicates the best fit line with its 1$\sigma$ confidence level, while the Spearman Rank coefficient, its uncertainty and statistical significance are reported in the top-left corner. \textit{Panel d).} Turn-off stars $\vb$ KDE distribution (azure line), together with the result of the GMM fit (grey line).} 
    \label{fig:ngc5822 toff}
\end{figure*}

The turn-off stars selection, illustrated in Fig.~\ref{fig:ngc5822 toff}a for the $\sim 1$ Gyr-old cluster NGC\,5822, have been carried out with the following criteria, adopting the best fit parameters from \citet{dias2021} for Parsec isochrones \citep{marigo2017}:
\begin{itemize}
    \item the faint magnitude limit (line BC) have been determined by eye as the magnitude where the color spread is notably larger than the MS width of fainter stars, and with the goal of excluding clear photometric binary star candidates, whose location is indicated by the black dashed line in Fig.~\ref{fig:ngc5822 toff}a\footnote{Unfortunately, due to the shape of the MSTO, it is not possible to exclude binary stars from the turn-off stars selected sample. Future spectroscopic studies will allow to better characterize the presence of binary stars among turn-off stars, thus allowing a detailed investigation of their rotation velocity distribution.};
    \item the bright limit (line DA) have been determined connecting the colors and magnitudes of the brightest MS stars from a sample of isochrones spaced by 0.10 in log age; 
    \item the red boundary (line CD) has been determined by eye, or, whenever feasible, as the red border of the color distribution\footnote{a good estimates of the color distribution border is possible only when the turn-off well populated};
    \item the blue boundary (line BA) has been determined in the same way as the red one, but for the blue side. 
\end{itemize} 

We then employed the blue and red fiducial lines to define the verticalized color as in \citet{milone2017a}, i.e. $\dc = w_\mathrm{TO}\frac{C - C_{\rm Blue}}{C_{\rm Red} - C_{\rm Blue}}$, where $C, C_\mathrm{Blue}, C_\mathrm{Red}$ represent the color of the star, and of the blue/red fiducial line, respectively, while $w_\mathrm{TO}$ is color width of the turn-off at a reference magnitude, i.e. approximately at 0.5 magnitudes brighter than the fainter limit BC. Consequently, stars on the blue and red fiducial lines correspond to $\dc=0$ and $\dc=w_\mathrm{TO}$, respectively. The turn-off reference magnitudes, together with the turn-off faint limits, are listed in Tab.~\ref{tab:tab1}. 

The verticalized CMD, i.e. $\rp$ vs. $\dc$, is shown in Fig.~\ref{fig:ngc5822 toff}b1, where stars with available $\vb$ have been colored according to the top colormap\footnote{It's worth noting that while a sequential colormap might be more appropriate given the range of $\vb$, i.e. from 0 to positive values only, we have chosen to use a blue and red colormap consistently with existing literature and exploratory studies on the eMSTO phenomenon. Historically, these colors have been associated with slow and fast rotators, respectively.}. The kernel density estimates (KDE) of $\dc$ for all selected turn-off stars is represented in Fig.~\ref{fig:ngc5822 toff}b2 by the azure line, and the 16$^{\rm th}$, 50$^{\rm th}$ and 84$^{\rm th}$ percentiles are indicated by the black vertical lines. To further describe the turn-off $\dc$ distribution, we exploited the \texttt{scikit-learn} python package \citep{scikit-learn} to use Gaussian Mixture Models (GMM) and fit $N$ number of Gaussians, with $N$ ranging from 1 to 3. The best-fit number of Gaussians has been determined as the one minimizing the Bayesian Information Criterion (BIC).

Unlike histograms and KDE, which depend on bin and bandwidth choice respectively, GMM is performed directly on the unbinned data distribution, making it independent of bin or bandwidth selection. Therefore, it offers a reliable and objective description of the underlying distribution. In the case of NGC\,5822, the two best fit Gaussians are represented in Fig.~\ref{fig:ngc5822 toff}b2 with the blue and red line respectively.

We investigate the relation between position in the turn-off, as indicated by $\dc$, and projected stellar rotation, i.e. $\vb$, in Fig.~\ref{fig:ngc5822 toff}c, where each point is color-coded according to its relative $\vb$ uncertainty as indicated in the top colorbar. Specifically, lighter colors indicate measurements with lower relative uncertainty. Furthermore, absolute errors on $\vb$ are indicated as gray errorbars. The red line and shaded region indicate the best-fit straight line, while the Spearman Rank coefficient, the associated uncertainty and the significance of the correlation are shown in top-left corner. We refer to Sec.~\ref{subsec:toff morphology} for a detailed description of how the uncertainty and significance of the correlation have been determined. 
Finally, the $\vb$ 1-dimensional KDE and the GMM fit of the $\vb$ distribution are shown as an azure and gray line in Fig.~\ref{fig:ngc5822 toff}d. 

In the following section, we describe how verticalized colors and projected stellar rotations have been exploited to describe different aspects of the turn-off morphology.

\subsection{Determining the turn-off morphology}\label{subsec:toff morphology}
    To quantify the color spread within the turn-off region, we compute the dispersion of the distribution, defined as the difference between the $16^\mathrm{th}$ and $84^\mathrm{th}$. In the following, we will refer to the color spread as to $S_{\dc}$. Furthermore, the uncertainties associated to $S_{\dc}$ have been determined by bootstrapping with replacement a 1000 times. 
    In the case of the $\sim 1$ Gyr cluster NGC\,5822, illustrated in Fig.~\ref{fig:ngc5822 toff}, we determine a color spread of $S_{\dc} = 0.070 \pm 0.006$. The results for the 32 clusters in the age range (50 Myr - 2.5 Gyr) will be presented and discussed in Sec.~\ref{sec:results}.
    
    To further describe the turn-off morphology, we computed the median $(\dc_{\rm Med})$ of the verticalized color distribution. This quantity was then normalized between 0 and 1 dividing it by the total color width of the turn-off color distribution. As a result, values close to 0 indicate a preference for stars to be located on the blue and bright side of the turn-off, while values of 1 suggest that the majority of stars tend to be distributed on the red and faint end of the turn-off. As an example, the location of the non-normalized median is shown in Fig.~\ref{fig:ngc5822 toff}b2 as the vertical line corresponding to the 50$^{\rm th}$ percentile, i.e. $\dc \sim 0.07$, while its normalized value corresponds to $\dc_\mathrm{Med}=0.355 \pm 0.017$.

    In order to explore the potential association between projected stellar rotation, indicated by $\vb$, and the position in the turn-off, quantified by $\dc$, we compute the Spearman Rank correlation coefficient $(R_{\rm S})$ of the $\vb$ vs. $\dc$ distribution. The value of $R_{\rm S}$ is presented in the top inset of Fig.~\ref{fig:ngc5822 toff}c, and its uncertainty is determined by means of bootstrap analysis.
    
    To assess the statistical significance of the Spearman rank coefficient, i.e. to determine if the observed correlation could be the result of random fluctuations and observational uncertainties, we adopted the following procedure. We generated 1000 samples with color distribution randomly drawn from the observed $\dc$ distribution and random uniform distribution in $\vb$ with minimum and maximum taken as from observed $\vb$. This step ensures that there is no initial correlation between the simulated $\vb$-$\dc$ distribution. To account for uncertainties, a random quantity drawn from a Gaussian with dispersion consistent with the observed uncertainties $\epsilon_{\vb}$ has been added to each simulated $\vb$. For each of the 1000 realizations we determined the correlation between simulated $\dc$ and $\vb$. We finally computed the significance of the observed $R_{\rm S}$, referred to as $p$, as the fraction of random realizations with correlation smaller than the observed value. Thus, values of $p$ approaching 0 suggest a high likelihood of uncertainties leading to correlations akin to the observed one, while values near unity indicate that the observed correlation is unlikely  the result of random fluctuations and uncertainties. 
    The Spearman coefficient, along with its uncertainty and the significance of the correlation for NGC\,5822, is presented in the top inset of Fig.~\ref{fig:ngc5822 toff}c. Specifically, we find a Spearman rank coefficient of $R_{\rm S}=0.592 \pm 0.159$, with a significance $p=0.999$, indicating that slow rotators are consistently found on the blue and bright side of the turn-off, while fast rotators are preferentially on the faint red side. Such result is consistent with the findings of \citet{marino2018a}, where the spectroscopic analysis of bright MS and turn-off stars in the Galactic Open cluster NGC\,6705 revealed a similar correlation between $\vsini$ and turn-off location. Similar results are also found in young Magellanic Clouds clusters with MUSE data \citep[see e.g.][]{kamann2020, kamann2023}

    Finally, to describe the distribution of projected rotational velocities for turn off stars, displayed in Fig.~\ref{fig:ngc5822 toff}d, we adopt a similar approach to the one described in the previous paragraphs. Specifically, we visually represent the distribution by means of Gaussian kernel KDE (azure line), and we perform a GMM analysis to determine the optimal number of Gaussians needed to fit the distribution (shown in gray in Fig.~\ref{fig:ngc5822 toff}d).

\section{Results}
\label{sec:results}

In this section we present and discuss the results obtained for the sample of analyzed clusters. Specifically, while the morphology of the eMSTO have been investigated for the whole sample of 32 OCs, due to the lower number of stars with available $\vb$, we could only explore the role of stellar rotation in 17 clusters. Clusters with available $\vb$ are flagged with the keyword \texttt{flag}$\vb$=\texttt{TRUE} in Table~\ref{tab:tab1}.   

\rowcolors{2}{gray!25}{white}
\begin{table*}
    \caption{Cluster properties and results. ID, cluster age right-ascension and declination from \citet{dias2021}, reference turn-off magnitude, faint turn-off magnitude limit, number of selected turn-off stars, flag for stellar rotation, Spearman rank coefficient and uncertainty, maximum measured $\vb$, color spread and median color, with uncertainties, double fiducial and single isochrone verticalization.}
    \centering
    \begin{adjustbox}{max width=\textheight, angle=90}
        \begin{tabular}{lcccccccccccccccccccc}
        \hline
        \hline
         Cluster & Age & RA & DEC & $\rp^\mathrm{TO}$ & $\rp^\mathrm{FAINT}$ & $N_\mathrm{stars}^\mathrm{TO}$ & \texttt{flag}$\vb$ & $R_\mathrm{S}$ & $\epsilon_\mathrm{R_\mathrm{S}}$ & $\vb^\mathrm{MAX}$ & $S_\mathrm{\dc^{fid}}$ & $\epsilon_\mathrm{S_\mathrm{\dc^\mathrm{fid}}}$ & $\dc_\mathrm{Med}^\mathrm{fid}$ & $\epsilon_\mathrm{\dc_\mathrm{Med}^\mathrm{fid}}$ & $S_\mathrm{\dc^{iso}}$ & $\epsilon_\mathrm{S_\mathrm{\dc^\mathrm{iso}}}$ & $\dc_\mathrm{Med}^\mathrm{iso}$ & $\epsilon_\mathrm{\dc_\mathrm{Med}^\mathrm{iso}}$ \\
         & Gyr & deg  & deg & mag & mag & log(yr) & log(yr) &  &  &  &  &  &  &  &  &  &  &  &  &  \\
        \hline
        \hline
        \textbf{Collinder\,394} & 0.086 & 283.060 & -20.224 & 8.400 & 9.900 & 12 & False & NaN & NaN & NaN & 0.014 & 0.003 & 0.272 & 0.050 & 0.031 & 0.005 & 0.021 & 0.012 \\
        \textbf{Haffner\,22}    & 2.438 & 123.104 & -27.902 & 14.400 & 14.900 & 101 & False & NaN & NaN & NaN & 0.017 & 0.001 & 0.367 & 0.010 & 0.021 & 0.002 & 0.045 & 0.002 \\
        \textbf{Haffner\,26}    & 0.587 & 123.906 & -30.848 & 13.900 & 14.400 & 41 & False & NaN & NaN & NaN & 0.068 & 0.009 & 0.429 & 0.018 & 0.068 & 0.007 & 0.086 & 0.016 \\
        \textbf{IC\,2714}       & 0.476 & 169.379 & -62.711 & 12.200 & 13.200 & 108 & False & NaN & NaN & NaN & 0.036 & 0.002 & 0.328 & 0.017 & 0.040 & 0.002 & 0.067 & 0.005 \\
        \textbf{Melotte\,20}    & 0.083 & 51.510 & 48.933 & 7 & 8.200 & 24 & True & 0.714 & 0.220 & 340.855 & 0.021 & 0.002 & 0.417 & 0.058 & 0.028 & 0.003 & 0.050 & 0.009 \\
        \textbf{Melotte\,22}    & 0.131 & 56.647 & 24.099 & 7 & 7.200 & 17 & True & 0.800 & 0.222 & 195.606 & 0.030 & 0.006 & 0.326 & 0.025 & 0.033 & 0.006 & 0.042 & 0.003 \\
        \textbf{Melotte\,71}    & 1.327 & 114.378 & -12.068 & 13.500 & 14 & 82 & False & NaN & NaN & NaN & 0.042 & 0.003 & 0.288 & 0.022 & 0.048 & 0.003 & 0.078 & 0.004 \\
        \textbf{NGC\,1245}      & 1.199 & 48.684 & 47.236 & 14.200 & 14.400 & 76 & False & NaN & NaN & NaN & 0.072 & 0.005 & 0.415 & 0.025 & 0.067 & 0.006 & 0.136 & 0.007 \\
        \textbf{NGC\,1664}      & 0.643 & 72.754 & 43.674 & 11.600 & 12.500 & 21 & True & 0.482 & 0.293 & 251.159 & 0.042 & 0.007 & 0.496 & 0.065 & 0.049 & 0.009 & 0.082 & 0.006 \\
        \textbf{NGC\,1817}      & 1.236 & 78.132 & 16.702 & 13 & 13.200 & 41 & False & NaN & NaN & NaN & 0.064 & 0.006 & 0.311 & 0.024 & 0.067 & 0.005 & 0.089 & 0.019 \\
        \textbf{NGC\,2099}      & 0.608 & 88.076 & 32.549 & 11.500 & 12.800 & 211 & True & 0.198 & 0.248 & 262.072 & 0.059 & 0.002 & 0.454 & 0.021 & 0.050 & 0.002 & 0.106 & 0.005 \\
        \textbf{NGC\,2281}      & 0.590 & 102.079 & 41.047 & 10 & 10.800 & 21 & True & 0.867 & 0.095 & 259.352 & 0.043 & 0.005 & 0.382 & 0.108 & 0.050 & 0.005 & 0.062 & 0.006 \\
        \textbf{NGC\,2287}      & 0.302 & 101.508 & -20.701 & 9.500 & 11.300 & 67 & True & 0.590 & 0.103 & 327.797 & 0.032 & 0.003 & 0.373 & 0.020 & 0.033 & 0.002 & 0.065 & 0.006 \\
        \textbf{NGC\,2323}      & 0.425 & 105.666 & -8.379 & 11.200 & 12 & 60 & False & NaN & NaN & NaN & 0.041 & 0.004 & 0.470 & 0.010 & 0.049 & 0.003 & 0.083 & 0.005 \\
        \textbf{NGC\,2358} & 0.578 & 109.246 & -17.117 & 11 & 11.700 & 8 & False & NaN & NaN & NaN & 0.029 & 0.004 & 0.311 & 0.081 & 0.036 & 0.011 & 0.039 & 0.004 \\
        \textbf{NGC\,2360} & 1.268 & 109.440 & -15.630 & 11.500 & 12.200 & 55 & True & 0.376 & 0.161 & 232.108 & 0.054 & 0.009 & 0.390 & 0.024 & 0.048 & 0.004 & 0.075 & 0.004 \\
        \textbf{NGC\,2420} & 2.223 & 114.605 & 21.572 & 14.200 & 14.400 & 54 & False & NaN & NaN & NaN & 0.008 & 0.001 & 0.329 & 0.006 & 0.012 & 0.001 & 0.033 & 0.001 \\
        \textbf{NGC\,2423} & 1.089 & 114.293 & -13.834 & 11 & 11.800 & 45 & True & 0.611 & 0.118 & 233.904 & 0.054 & 0.006 & 0.414 & 0.013 & 0.048 & 0.006 & 0.093 & 0.005 \\
        \textbf{NGC\,2447} & 0.668 & 116.157 & -23.866 & 10.500 & 12 & 67 & True & 0.516 & 0.107 & 345.945 & 0.042 & 0.004 & 0.328 & 0.040 & 0.041 & 0.004 & 0.051 & 0.007 \\
        \textbf{NGC\,2527} & 0.800 & 121.237 & -28.099 & 10.350 & 11.200 & 36 & True & 0.612 & 0.141 & 231.426 & 0.050 & 0.006 & 0.476 & 0.050 & 0.047 & 0.003 & 0.069 & 0.003 \\
        \textbf{NGC\,2548} & 0.521 & 123.400 & -5.719 & 10.500 & 11.500 & 17 & True & 0.767 & 0.193 & 290.212 & 0.038 & 0.006 & 0.185 & 0.019 & 0.050 & 0.007 & 0.054 & 0.006 \\
        \textbf{NGC\,2632} & 0.762 & 130.057 & 19.613 & 8 & 8.300 & 24 & True & 0.117 & 0.491 & 214.738 & 0.058 & 0.007 & 0.532 & 0.065 & 0.047 & 0.006 & 0.066 & 0.007 \\
        \textbf{NGC\,2658} & 0.540 & 130.865 & -32.662 & 14 & 14.400 & 38 & False & NaN & NaN & NaN & 0.054 & 0.006 & 0.336 & 0.049 & 0.047 & 0.006 & 0.093 & 0.006 \\
        \textbf{NGC\,2818} & 1.167 & 139.045 & -36.621 & 13.900 & 14.400 & 33 & False & NaN & NaN & NaN & 0.069 & 0.008 & 0.403 & 0.021 & 0.072 & 0.012 & 0.089 & 0.014 \\
        \textbf{NGC\,3114} & 0.228 & 150.572 & -60.056 & 10 & 10.500 & 46 & True & 0.636 & 0.252 & 314.674 & 0.029 & 0.003 & 0.316 & 0.021 & 0.036 & 0.003 & 0.055 & 0.005 \\
        \textbf{NGC\,3532} & 0.413 & 166.411 & -58.722 & 8.500 & 10.300 & 150 & True & 0.569 & 0.120 & 296.500 & 0.051 & 0.002 & 0.370 & 0.024 & 0.041 & 0.001 & 0.071 & 0.003 \\
        \textbf{NGC\,5822} & 1.094 & 226.085 & -54.365 & 11 & 11.300 & 51 & True & 0.592 & 0.159 & 257.757 & 0.070 & 0.006 & 0.355 & 0.017 & 0.070 & 0.008 & 0.143 & 0.007 \\
        \textbf{NGC\,6208} & 1.914 & 252.327 & -53.713 & 12.400 & 13 & 41 & False & NaN & NaN & NaN & 0.052 & 0.006 & 0.434 & 0.061 & 0.052 & 0.005 & 0.100 & 0.011 \\
        \textbf{NGC\,6705} & 0.294 & 282.765 & -6.269 & 12.400 & 12.500 & 113 & False & NaN & NaN & NaN & 0.067 & 0.006 & 0.538 & 0.019 & 0.060 & 0.003 & 0.112 & 0.003 \\
        \textbf{NGC\,6940} & 1.175 & 308.615 & 28.273 & 12 & 12 & 47 & True & 0.401 & 0.130 & 224.505 & 0.067 & 0.005 & 0.623 & 0.029 & 0.058 & 0.004 & 0.115 & 0.004 \\
        \textbf{NGC\,7209} & 0.494 & 331.229 & 46.495 & 11.500 & 13 & 65 & True & 0.727 & 0.076 & 309.506 & 0.062 & 0.005 & 0.415 & 0.038 & 0.047 & 0.004 & 0.105 & 0.009 \\
        \textbf{Trumpler\,29} & 0.053 & 265.359 & -40.169 & 11 & 12.500 & 16 & False & NaN & NaN & NaN & 0.013 & 0.002 & 0.208 & 0.064 & 0.022 & 0.002 & 0.005 & 0.009 \\
        \hline
        \hline
        \end{tabular}
    \end{adjustbox}
\label{tab:tab1}
\end{table*}

\begin{figure*}
    \centering
    \includegraphics[width=0.99\textwidth, trim={0cm 0cm 0cm 0cm}, clip]{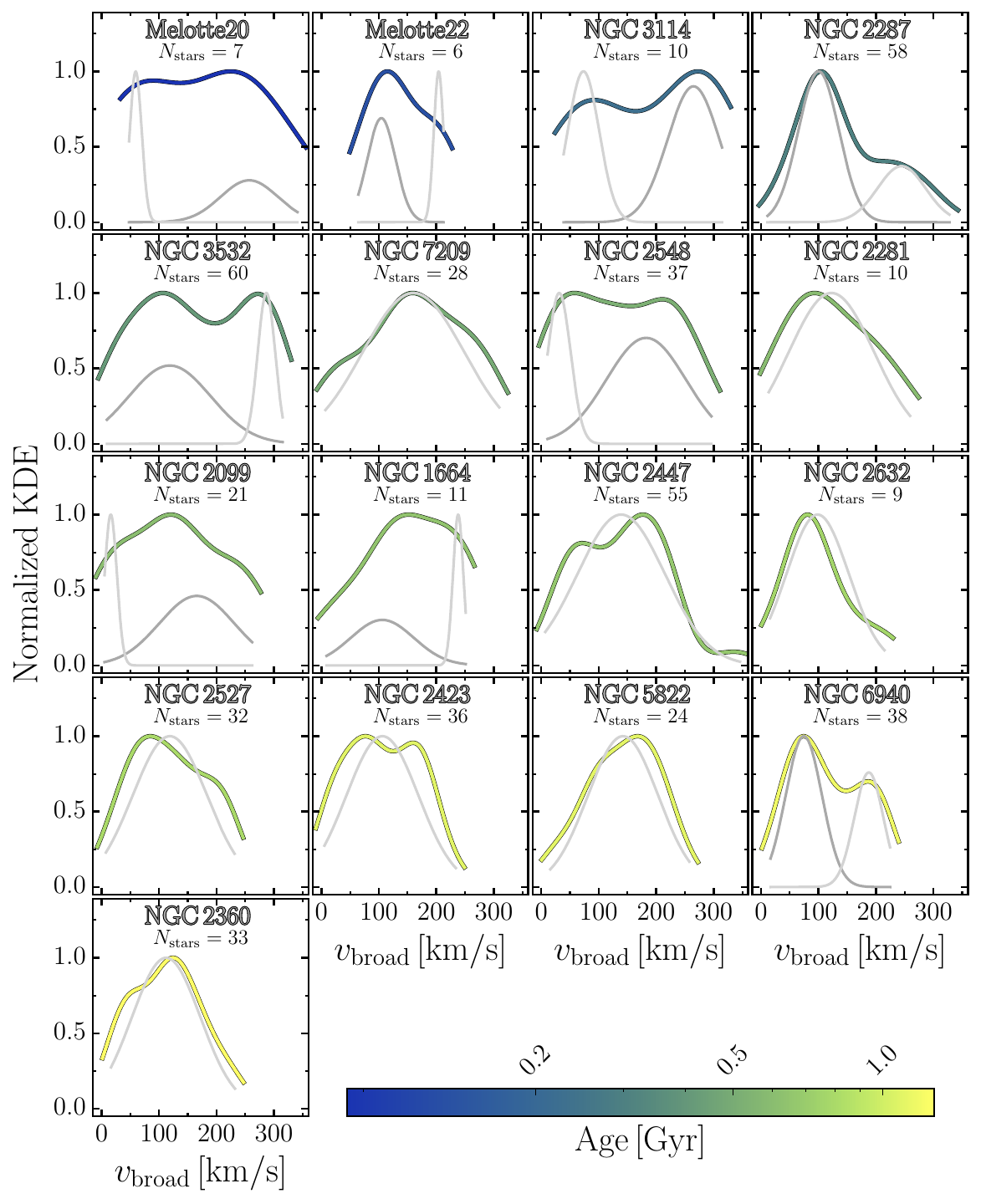}
    \caption{$\vb$ Kernel density estimate (KDE) distribution (colored lines) and Gaussian Mixture Model fit (gray lines) for the 17 analyzed clusters. Clusters are sorted by age, and each KDE is colored according to cluster age as shown in the bottom colorbar.} 
    \label{fig:allvb dist}
\end{figure*}

We refer to Sec.~\ref{sec:toff} for a detailed description of how the different quantities, e.g. $\dc, \, S_{\rm \dc}, \, \dc_{\rm Med}$ etc. have been determined. Moreover, while we provide the information on the $\vb$ distribution of turn-off stars, we refrain from carrying out a detailed and quantitative analysis in each individual cluster. Indeed, considering the small number of stars in some clusters, as well as the fact that Gaia $\vb$ selection function is not exactly known, it would be impossible to quantify and correct for possible observational biases. Nonetheless, we find worth showing the rotational velocity distribution of turn-off stars for all the 17 clusters in Fig.~\ref{fig:allvb dist}. The description  of the procedure used to derive the distributions, as well the GMM fit is described in Sec.~\ref{subsec:toff morphology}. The KDE distributions are shown with solid colored lines, as indicated by the bottom colorbar, while the light and dark gray lines indicate the different best-fit Gaussians returned by the GMM fit. We point out that that, while the KDE/histograms are sensitive to the bandwidth/bin choice, the GMM fit is performed directly on the unbinned $\vb$ values, so that there is no direct connection between the best-fit Gaussians and the KDE/histograms in Fig.~\ref{fig:allvb dist}. Nonetheless, as in some clusters the fit is performed on a low number of stars, we do not take the results of the GMM as a quantitative results. Clusters are sorted by age, and the number of stars from which the distribution is determined, namely $N_\mathrm{stars}$, is listed in the top inset.

We find that 7 out of 17 clusters exhibit $\vb$ distribution that are best reproduced by two Gaussians, while the remaining 10 are consistent with a distribution  drawn from a single Gaussian. Remarkably, the bimodality seems to be more common among younger stars clusters, with the four youngest clusters all exhibiting bimodal distributions. Therefore, although the low number of stars and the unknown Gaia $\vb$ selection function prevent us from drawing firm conclusion, we point out that the $\vb$ distribution of younger clusters is qualitatively consistent with an approximately bimodal distribution, as found for Magellanic Clouds clusters \citep[see e.g.][]{marino2018a, kamann2020, kamann2023}. Nonetheless, we avoid trying to quantitatively explain the observed individual distributions.

\subsection{Turn-off morphology as a function of cluster age} \label{subsec: dcage}
In the last decades, evidence of the pivotal role of stellar rotation in the shaping and evolution of young star clusters are revolutionizing our understanding of these -unexpectedly- complex objects. 

\begin{figure*}
    \centering
    \includegraphics[width=0.99\textwidth, trim={0cm 0cm 0cm 0cm}, clip]{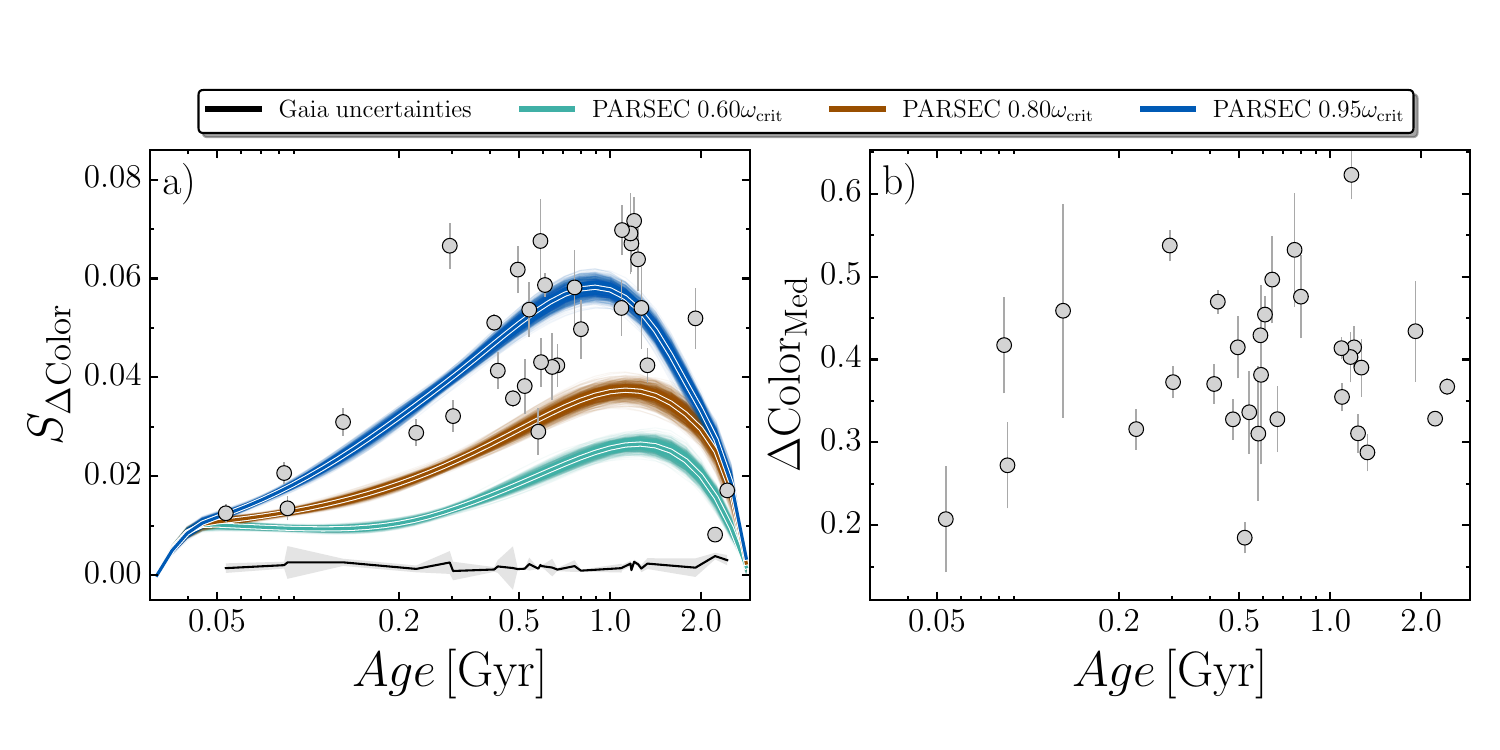}
    \includegraphics[width=0.49\textwidth, trim={0cm 0cm 0cm 0cm}, clip]{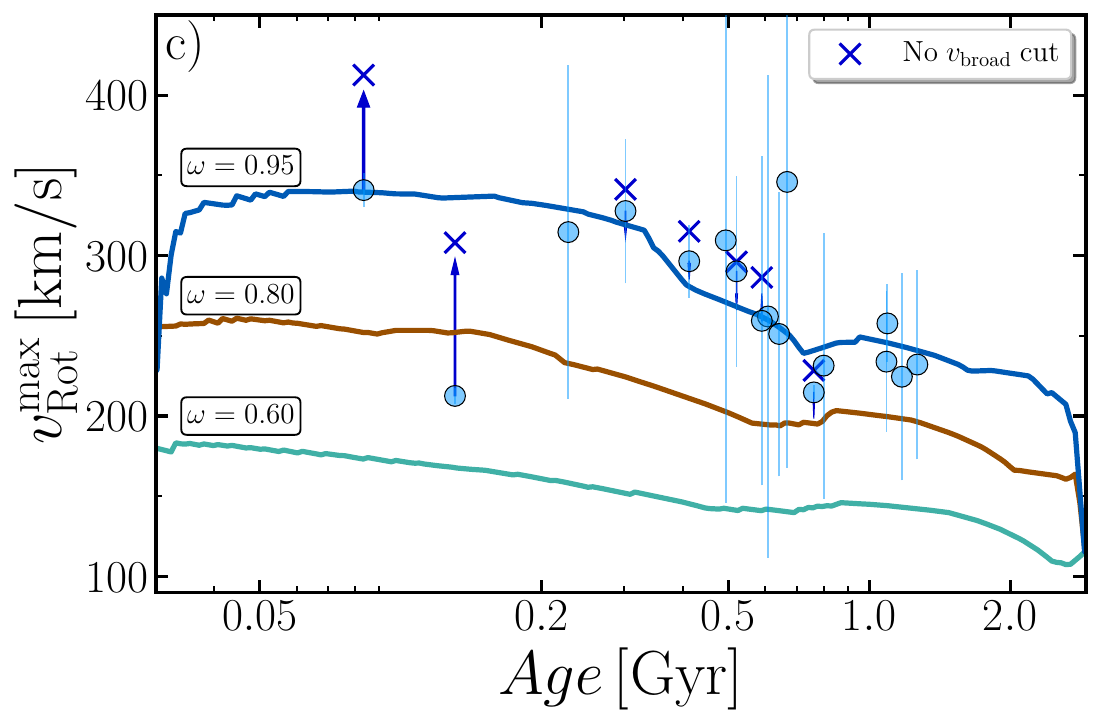}
    \caption{\textit{Panel a).} Normalized color spread, determined as discussed in Sec.~\ref{subsec:toff morphology}, vs. cluster age. Uncertainties are shown as gray errorbars. Green, red and blue lines indicate the cubic spline fit of the predictions from Parsec models with rotation velocities of respectively 0.60, 0.80, 0.95\,$\omega_\mathrm{crit}$, while the uncertainties on the predicted relations are shown as shaded thin lines. Gaia photometric uncertainties are shown as black solid line in the bottom of the panel. \textit{Panel b).} Peak of the normalized color distribution vs. cluster age. \textit{Panel c).} Comparison between maximum observed and expected rotational velocities as a function of cluster for models with different initial stellar rotation. Predictions from models with different rotation are shown with colored lines, while azure dots indicate the observations for Galactic OCs.} 
    \label{fig:col dist}
\end{figure*}

In this work, we homogeneously analyze the turn-off morphology of 32 clusters younger than 2.4 Gyr by means of Gaia DR3 astrometry and photometry. Figure~\ref{fig:col dist} presents the color spread ($S_\mathrm{\dc}$, panel a) and normalized median color ($\dc_{\rm Med}$, panel b) in relation to cluster age. Uncertainties on both quantities have been determined by bootstrapping with replacement the results.

A visual inspection of Fig.~\ref{fig:col dist}a1 reveals a clear correlation between color spread and cluster age for clusters younger than 2 Gyr, while in older clusters the observed color spread drastically decreases to values approximately consistent with those expected from Gaia photometric uncertainties, depicted as a black line with shaded 1$\sigma$ region in the bottom of Fig.~\ref{fig:col dist}a. Remarkably, the two oldest clusters in our sample, namely Haffner\,22 ($Age = 2.44\,{\rm Gyr}$) and NGC\,2420 ($Age = 2.22\,{\rm Gyr}$), exhibit a color spread smaller than that of the youngest cluster, i.e. Trumpler\,29 ($Age = 50\,{\rm Myr}$). 

We then compared the observed $\dc$ vs. age distribution with the theoretical predictions derived from Parsec isochrones with different rotational velocities \citep{marigo2017, girardi2019, nguyen2022}. More details about rotating Parsec isochrones are provided in Appendix~\ref{app:parsec}.

Color spread-age relations for different rotational velocities are shown with colored lines, as indicated in the top legend. For clarity, in Fig.~\ref{fig:col dist}a we only show the results for $\omega=0.60, 0.80, 0.95\omega_{\mathrm{crit}}$. We refer to \citet{girardi2019, nguyen2022} for a detailed treatment of the physical prescriptions adopted to compute the rotating isochrones. 
However, we mention here two key details: 
\begin{enumerate}
    \item while each star in the isochrone starts its life with the reference initial rotation (e.g. $\omega=0.6, \dots, 0.99\,\omega_\mathrm{crit}$, for the masses having radiative envelopes), the evolution changes the surface rotation so that different masses exhibit different rotation rates at different epochs and evolutionary stage.
    As a consequence, each isochrone and simulation intrinsically incorporates stars with different rotation rates, from 0 for the lower mass stars, to a maximum value decreasing for increasing age.
    \item Different stellar models adopt different definition of critical rotation velocity. Specifically, Parsec and Geneva stellar models \citep{ekstrom2012, georgy2013}  define $\omega_\mathrm{crit} = (2/3)^{3/2} \sqrt{GM/R^3_\mathrm{pol}}$ while in MESA models \citep{dotter2016, choi2016} $\omega_\mathrm{crit} = \sqrt{GM/R^3_\mathrm{e}}$, where $R_\mathrm{pol}, R_\mathrm{e}$ indicate the polar and equatorial radii. As a consequence, $\omega = 0.9 \omega_\mathrm{crit}$ in Parsec/Geneva models corresponds to $\omega = 0.5 \omega_\mathrm{crit}$ in MESA models\citep[see e.g][]{wang2023}.
\end{enumerate}  

In the following, we use the definition of Parsec isochrones.

To determine the predictions for different initial rotation rates, we first created a population of 100,000 stars, drawing masses from a Salpeter initial mass function (IMF) and inclinations of rotation axes from an isotropic distribution ranging between 0 and 90$^{\circ}$. Using the mass-inclination-magnitude relationship provided by Parsec isochrones at a specific age, we then determined the $G, \bp, \rp$ magnitudes for each couple of mass and inclination. 
Simulations were generated from approximately 30 Myr to around 2.5 Gyr, spaced by 0.1 in log age, and the trends displayed in Fig.~\ref{fig:col dist}a represent the cubic spline fit of the simulated $S_\mathrm{\dc}$. Uncertainties associated with the theoretical predictions, shown with colored shaded lines, were determined through 1000 iterations of the fitting process, where, at every iteration, each point was drawn from a Gaussian distribution centered on the simulated $\dc$ and with a dispersion equal to its uncertainty. Clearly, there is a very good agreement between the observed color spreads and the predictions from $\omega=0.95\omega_\mathrm{crit}$ models, while models with lower rotation rates cannot fully reproduce the observed color spread. This result is consistent with the findings of \citet{cordoni2018} and \citet{niederhofer2015} for smaller samples of OCs and Magellanic Clouds clusters, respectively

Furthermore, the distribution of the normalized median color of the turn-off stars, shown in Fig.~\ref{fig:col dist}b, exhibit hints of a mild trend with cluster age for clusters younger than 1 Gyr and becomes flat for older clusters. Such result is qualitatively consistent with the findings of \citet{milone2023a}, even though their correlation extends to all clusters younger than 2 Gyr. Similar conclusions are found when adopting a different color verticalization procedure.

Additionally, we compared the maximum measured projected rotational velocities in each cluster, denoted as $v_\mathrm{broad}^\mathrm{MAX}$, with the maximum rotational velocities predicted by Parsec isochrones as a function of cluster age. This comparison is shown in Fig.~\ref{fig:col dist}c, where predictions from models with different initial rotations are represented by colored lines, as indicated in the top colorbar. Our analysis suggests that Galactic open clusters likely host stars that formed as fast rotators, i.e. $\omega > 0.9 \omega_\mathrm{crit}$. The exception is the cluster Melotte\,22, whose maximum projected rotational velocity is lower than expected from fast-rotating models. However, this cluster does include faster rotators that align with the general trend but were excluded based on criteria discussed in \citet{fremat2023}. \\
Remarkably, Parsec models also accurately reproduce the changes in slope at around 0.4 and 0.9 Gyr. We compared the observed $v^\mathrm{MAX}_\mathrm{broad}$ vs. age relation with predictions from the Geneva models \citep{georgy2013} and found consistent results. Unfortunately, since public rotating MESA isochrones do not provide information about stellar rotation across the isochrones, we could not make the same comparison with MESA models.\\
Therefore, based on our analysis of the CMD and projected stellar rotation, we conclude that Galactic open clusters are consistent with the presence of stars that originally formed as fast rotators, i.e., $\omega \gtrsim 0.9 \omega_\mathrm{crit}$, with $\omega_\mathrm{crit}$ defined in Parsec models.

\subsection{Stellar rotation across the turn-off}
Stars in young open clusters exhibit projected rotational velocities ranging from few km/s to values as large as $\sim 300$ km/s, as shown in Fig.~\ref{fig:allvb dist}. In the following we will investigate the connection between projected stellar rotational velocities, as indicated by $\vb$, and MSTO morphology. 

\begin{figure*}
    \centering
    \includegraphics[width=0.99\textwidth, trim={0cm 0cm 0cm 0cm}, clip]{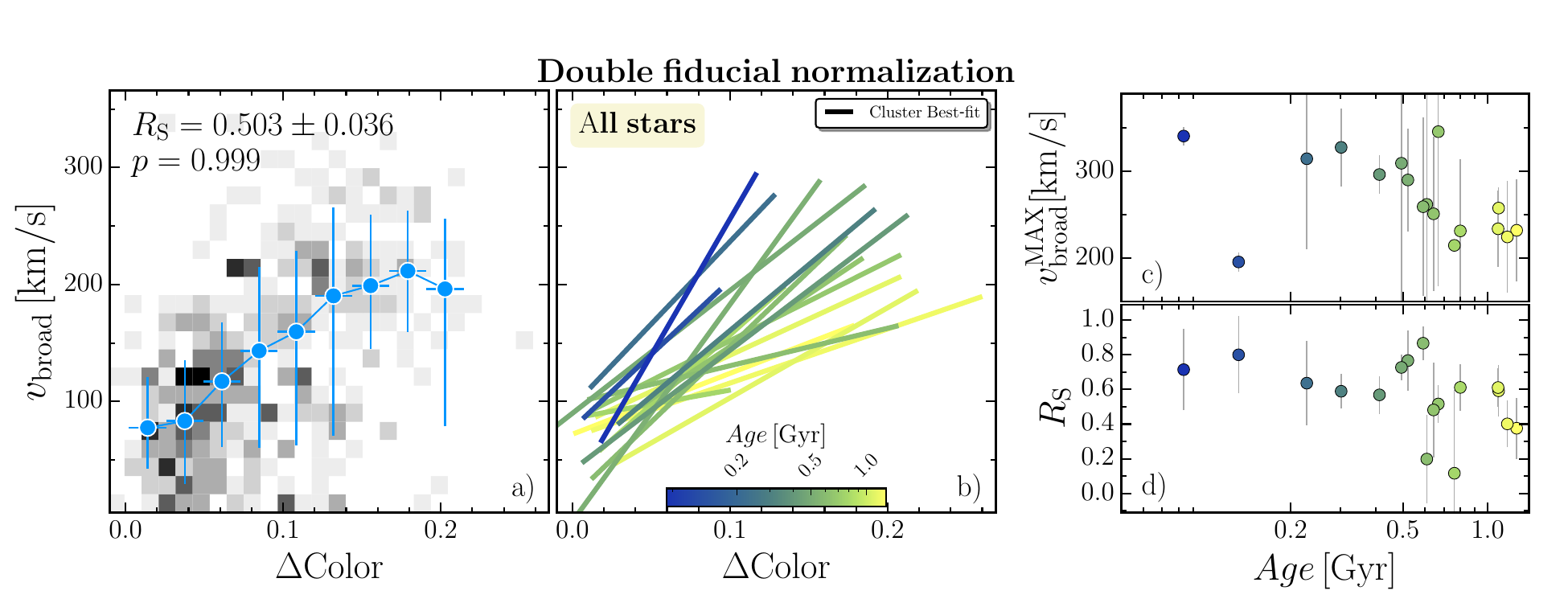}
    \caption{\textit{Panel a).} $\vb$ vs. $\dc$ of selected turn-off stars in all 17 clusters. Gray scale is indicative of the number of stars in each bin, from white to black. Azure points and errorbars represent the median trend with its uncertainties, computed in the equal width bins. The Spearman rank coefficient, with its uncertainty and significance is shown in the top-left inset. \textit{Panel b).} Best-fit line of $\vb$ vs. $\dc$ for each cluster, color-coded according to cluster age, as depicted in the bottom colorbar. \textit{Panel c).} Maximum observed $\vb$ of turn-off stars, with uncertainties as provided by Gaia DR3. \textit{Panel d).} Clusters' $\vb$-$\dc$  Spearman rank coefficient, $R_{\rm S}$, vs. clusters age. The uncertainties have been determined by bootstrapping the results, as discussed in the text.} 
    \label{fig:dcol vb}
\end{figure*}

Figure~\ref{fig:dcol vb} displays $\vb$ and $\dc$ for the 17 clusters where rotation velocities were available for a sufficient number of stars. Specifically, Fig.~\ref{fig:dcol vb}a shows the 2d histograms of all stars with available $\vb$ vs. $\dc$, with the median trend indicated by the azure points and lines. The azure errorbars represent the dispersion in each bin.  
Moreover, Fig.~\ref{fig:dcol vb}a illustrates a general positive correlation between projected stellar rotational velocity and color in the turn-off region. In other words, stars situated on the blue side of the turn-off tend to exhibit lower projected rotational velocities, while those on the red end of the turn-off are more likely to be fast rotators. The overall Spearman rank coefficient is $R_{\rm S} = 0.503 \pm 0.036$. By using the procedure described in Sec.~\ref{sec:toff}, we computed the statistical significance of the observed correlation, and found that the observed correlation is significant beyond the 3$\sigma$ level. Furthermore, when investigating the $\vb$-$\dc$ distribution of each cluster individually, whose best-fit line is depicted in Fig.\ref{fig:dcol vb}b, a dependence on cluster age arises. Specifically, younger clusters (marked with darker colors) exhibit a higher correlation than their older counterparts. This conclusion is supported by both the steeper best-fit line (Fig.~\ref{fig:dcol vb}b) and the higher Spearman rank coefficient (Fig.~\ref{fig:dcol vb}d) of younger clusters. The relation between $\vb$-$\dc$ correlation and cluster age is a consequence of two different mechanisms: \textit{i)} the maximum projected rotational velocity decreases with cluster age, as can be seen in Fig.~\ref{fig:dcol vb}c, and \textit{ii)} the turn-off color spread increases with cluster age (Fig.~\ref{fig:col dist}a). The combined effect is that as stars get older, they tend to occupy the lower $\vb$ and larger $\dc$ region, i.e. toward the bottom right corner of the $\vb$-$\dc$ plane.

Moreover, to assess the impact of Gaia $\vb$ observational uncertainties we repeated the analysis adopting different quality selections on $\vb$. Specifically, we selected stars with$\epsilon_{v_{\rm broad}}/v_{\rm broad} < 0.10, 0.25, 0.50, 1.00$. We find that the overall correlation between projected rotational velocities and normalized colors, as well as the the relation between correlation and cluster age, strengthen for stars with lower relative $\vb$ uncertainties. Such results and conclusions are also true when adopting a different verticalization process.

\begin{figure}
    \centering
    \includegraphics[width=0.5\textwidth, trim={0cm 0cm 0cm 0cm}, clip]{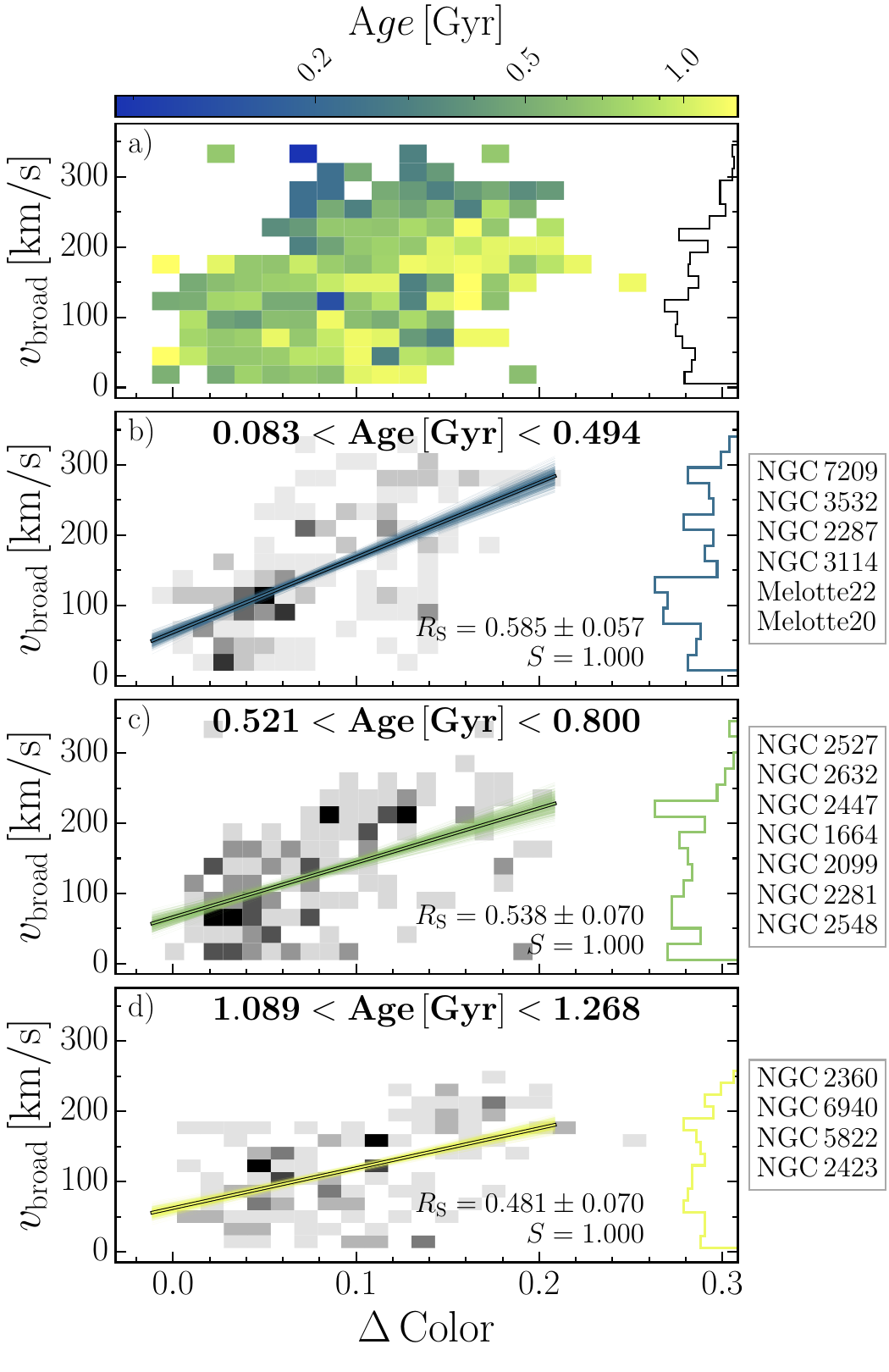}
    \caption{\textit{Panel a).} $\vb$ vs. $\dc$ of selected turn-off stars in all 17 clusters as in Fig.~\ref{fig:dcol vb}a, where each bin has been color-coded according to the mean age of stars in the bin, as indicated by the top colorbar. Specifically, darker colors indicate younger ages, while yellow-ish colors represent older ages.   \textit{Panel b-d).} Same as Fig.~\ref{fig:dcol vb}a with clusters divided into three age groups, as indicated in the top inset. The list of clusters falling in each specific bin is displayed in the right insets. Colored solid and shaded lines represent the best-fit line and 1000 Monte Carlo realizations, while the $\vb$ distribution of stars in each group is represented by the histograms on the right. The correlation, uncertainty and significance for each group is shown in the bottom-right corner.} 
    \label{fig:dcol vb age}
\end{figure}

To further explore and confirm the possible connection between $\vb$-$\dc$ correlation and cluster age, we divided the sample of analyzed clusters into three groups based on their age and repeated the analysis. Fig.~\ref{fig:dcol vb age}a shows the overall $\vb$-$\dc$ distribution, where each bin has been colored according to the mean age of stars in the specific bin. Remarkably, a visual inspection of Fig.\ref{fig:dcol vb age}a hints at the presence of a mild age gradient, with faster rotators generally belonging to younger star clusters. Additionally, redder stars, i.e. stars with higher $\dc$ values, tend to be older (i.e. yellow-ish colors). This pattern is clearly highlighted in Fig.~\ref{fig:dcol vb age}b-c, where we show the $\vb$-$\dc$ distributions for stars in the three age groups (youngest in the top panel, oldest in the bottom one). Specifically, younger stars display a slightly higher correlation between $\vb$ and $\dc$, as indicated by higher $R_{\rm S}$ values ($R_{\rm S} = 0.585 \pm 0.057$) and a steeper best-fit line. On the other hand, older clusters exhibit a lower correlation ($R_{\rm S} = 0.538 \pm 0.070$ and $R_{\rm S} = 0.481 \pm 0.070$) and a flatter relation. Correlations for all three subsets is significant beyond the 3$\sigma$ level.

Finally, the $\vb$ histograms on the right of each panel show that younger clusters host faster-rotating stars, with projected rotational velocities as high as 330 km/s. Moreover, concerning young star clusters, the distribution of rotational velocity distribution exhibits hints of bimodality, similarly to what is observed in Fig.~\ref{fig:allvb dist} and in young Magellanic Clouds clusters \citep[see e.g. ][]{marino2018a, kamann2020, kamann2023}.

Overall, regardless of $\vb$ observational uncertainties and/or color verticalization procedure, our analysis strongly supports the pivotal role of stellar rotation in the shaping of the turn-off in young star clusters. Furthermore, our results highlight a clear connection between stellar rotation and cluster age, hinting at a common evolutionary phenomenon.

\section{Discussion and interpretation}\label{sec:discussion}
In the following section we provide a tentative interpretation of the observed properties of eMSTOs in young Open clusters. Specifically, we exploit Parsec stellar models \citep{marigo2017, girardi2019, nguyen2022} to investigate and interpret the observed features of the analyzed Galactic OCs.   

To understand the observed correlation between stellar rotation and color, we exploit Parsec isochrones with different initial rotation rates and rotation axis inclination. 

\begin{figure*}
    \centering
    \includegraphics[width=0.49\textwidth, trim={0cm 0cm 0cm 0cm}, clip]{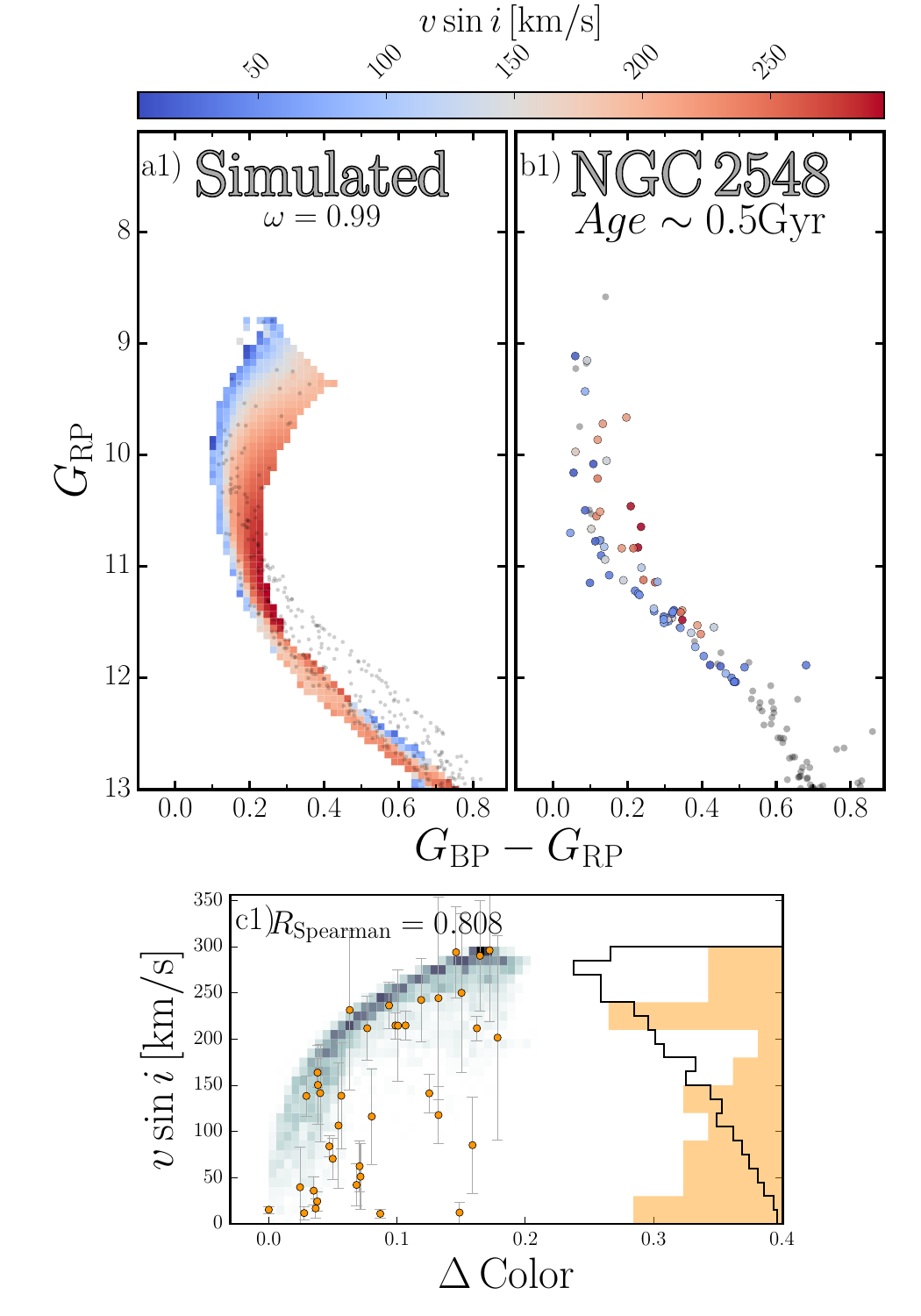}
    \includegraphics[width=0.49\textwidth, trim={0cm 0cm 0cm 0cm}, clip]{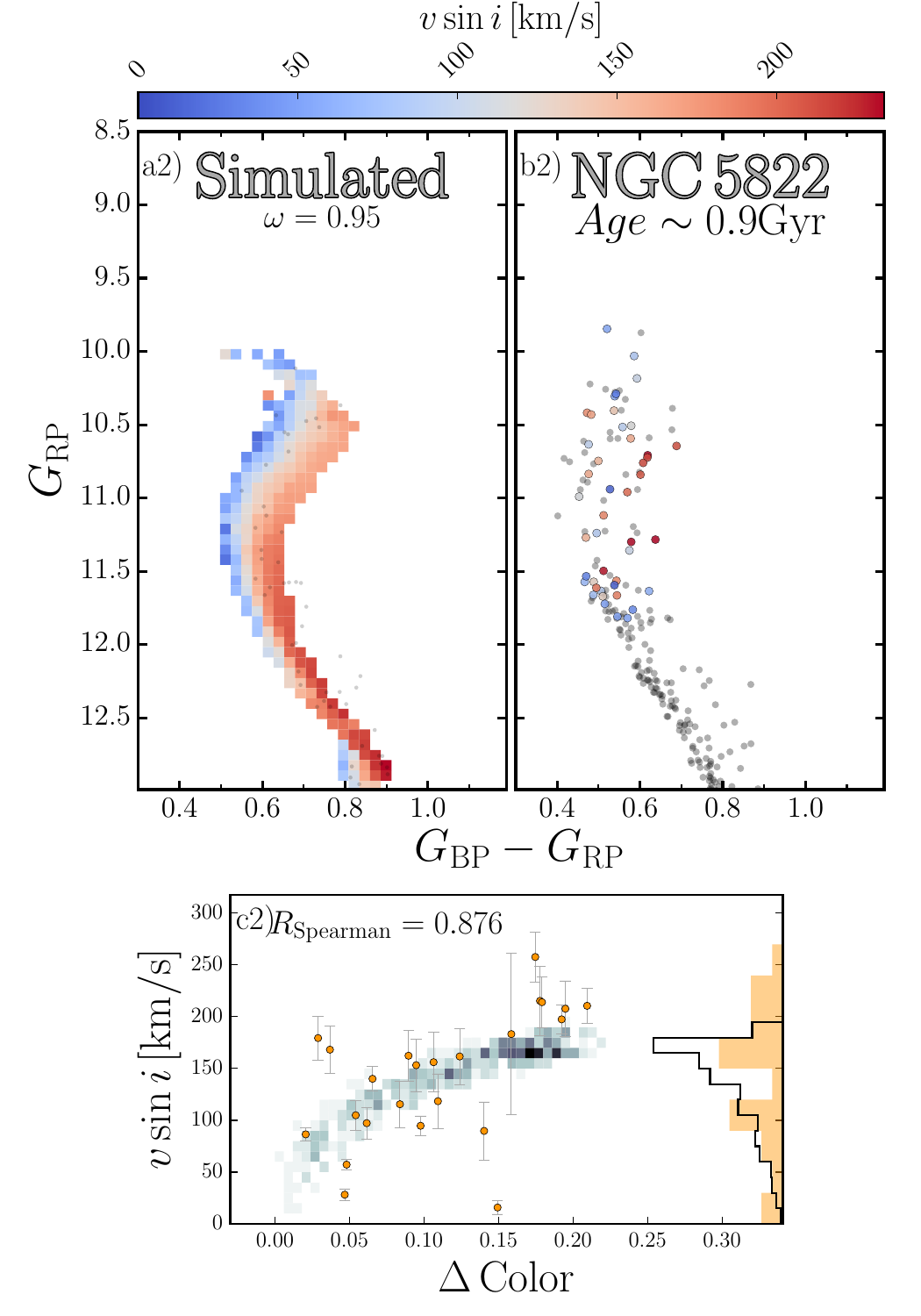}
    \caption{Comparison between theoretical simulations and the CMDs of NGC\,2548 (left panels) and NGC\,5822 (right panels) \textit{Panels a).} Simulated clusters with ages similar to NGC\,2548 and NGC\,5822 composed of a stellar population with rotation velocity $\omega=0,99, 0.95\omega_{\rm crit}$ and isotropic rotation axis inclination distribution. The color of each bin is indicative of the mean $\vsini$ of stars in the bin, as indicated by the top colorbar. Black dots represent binary stars. \textit{Panels b).} Observed CMDs with stars color-coded according to the observed Gaia $\vb$. \textit{Panels c).} Comparison between the observed $\vb$ vs $\dc$ (orange points) and the results of the simulation (gray scale). The normalized colors of the simulated stars have been computed in the same way as for observed stars, described in Sec.~\ref{subsec:toff morphology}. Orange shaded and black histograms represent the observed and simulated turn-off $\vb$ distribution.}
    \label{fig:sim}
\end{figure*}

In the following, we use two clusters, namely NGC\,2548 ($\sim 0.5\,\mathrm{Gyr}$) NGC\,5822 ($\sim 1\,\mathrm{Gyr}$), as comparison cases. As first step, we simulated $N=100,000$ stars with masses drawn from a Salpeter mass function, inclination of the stellar rotation axis following an isotropic distribution between 0 and 90 degrees, as found in \citet{healy2023}, and $\sim 17-30$\% of unresolved binary stars, as determined in \citet{cordoni2023}\footnote{Unresolved MS-MS binaries have been added to display their location and contribution to the main-sequence turn-off. However, considering that the rotational velocity distribution of binary stars is still poorly constrained, we avoid artificially assigning rotational velocities to binary stars.}. Moreover, a star seen pole-on, i.e. $i=0^\circ$, will have $\vsini=0$, while a stars observed edge-on, i.e. $i=90^\circ$, will have $\vsini=v_\mathrm{tangential}$. The choice of an isotropic inclination distribution is motivated by the recent findings of \citet{healy2023}, who, by exploiting spectroscopic and photometric data, determined that 8 out of 10 analyzed OCs exhibit rotation-axis orientations consistent with isotropy \citep[see also][for a detailed desription of the adopted procedure]{healy2020, healy2021}. A notable exception is the cluster NGC\,2548, which displays a certain degree of spin alignment and an apparently bimodal inclination distribution \citep[see e.g. Fig.~4 and 9 and Sec.~5.7.1 of][for a detailed description]{healy2023}. While our procedure is not accurate enough to discern between isotropy and/or alignment in the rotation axis distribution, we find worth noting that NGC\,2548 exhibits a bimodal rotation distribution (see e.g. Fig.~\ref{fig:allvb dist} and \ref{fig:sim}c1) and a strong correlation between projected rotation and color ($R_\mathrm{S}=0.767 \pm 0.193$). This results would be consistent with the presence of a single high initial rotation rate coupled with a bimodal inclination distribution. 

We then exploited Parsec isochrones with different rotation rates, namely $0.00, 0.30, 0.60, 0.80, 0.90, 0.95, 0.99\, \omega_{\rm crit}$, to generate synthetic photometry in the Gaia photometric system from the simulated masses and inclinations. Finally, we created the simulated CMD by adding photometric uncertainties. Two example simulated CMDs are depicted in Fig.~\ref{fig:sim}, where we adopted a distance modulus, reddening and age values \citep[as derived in][]{dias2021} consistent with NGC\,2548 (left) and NGC\,5822 (right). The synthetic clusters are composed of stars with initial rotation velocity equal to $\omega = 0.99 \omega_{\rm crit}$ and 0.95, respectively for the left and right simulations. Finally, we followed the procedure described in Sec.~\ref{subsec:toff morphology} to verticalize the color of turn-off stars and investigate the relation between $\vsini$ and $\dc$, with the results shown as gray-black 2D histograms. Orange points indicate observed NGC\,2548 and NGC\,5822 stars.

Comparing the simulated and observed CMDs (Fig.~\ref{fig:sim}a1-b1 and a2-b2) it can be noted that the turn-off and MS color spread, and overall CMD properties of both clusters are well reproduced by the simulations. 
Additionally, there is a good qualitative agreement between the simulations and observations, both in terms of recovered correlations ($R_\mathrm{S}=0.767 \pm  0.193$ for NGC\,2548 and $R_{\rm S}=0.592 \pm 0.159$ for NGC\,5822), and overall $\vsini$-$\dc$ distribution.\\
Similarly to the results presented in Sec.~\ref{subsec: dcage}, high initial rotation rates and inclinations ranging from 0 to 90$^\circ$, are mandatory to match the observed projected rotational velocity range and overall CMD properties. Moreover, to reproduce simultaneously the turn-off color spread and the width of the main-sequence, all stars must have similar initial rotation rate. Indeed, \textit{i)} a distribution of initial rotation rates would produce a considerable spread at the level of the main-sequence which is not observed in the CMDs of Open clusters; and \textit{ii)} a narrower range in inclination would produce a lower turn-off color spread. Both effects can be observed in the theoretical isochrones of Fig.~\ref{fig:iso parsec}. 
On the other hand, the simulated $\vsini$ distributions shown on the right of panels c1-c2 as black histograms, are not consistent with the observed $\vb$ distributions (shaded orange histograms). The observed projected rotation distributions could be reproduced by modeling an ad-hoc inclination distribution. However, we avoid carrying out a quantitative comparison. Indeed, while a detailed comparison between theoretical models and observations may provide important insights, there are too many unknowns to carry out a quantitative comparison, e.g. influence of binary stars, stellar mergers and Be-stars \citep{milone2018, kamann2021, wang2022, wang2023, bu2024}, Gaia $\vb$ completeness, magnetic and/or tidal braking effects \citep{dantona2017}, UV-dim stars and the possible influence of circumstellar dust \citep{milone2023a, martocchia2023, dantona2023}, and possible bimodal initial rotation rate \citep{bastian2018}.\\

\begin{figure}
    \centering
    \includegraphics[width=0.49\textwidth, trim={0cm 0cm 0cm 0cm}, clip]{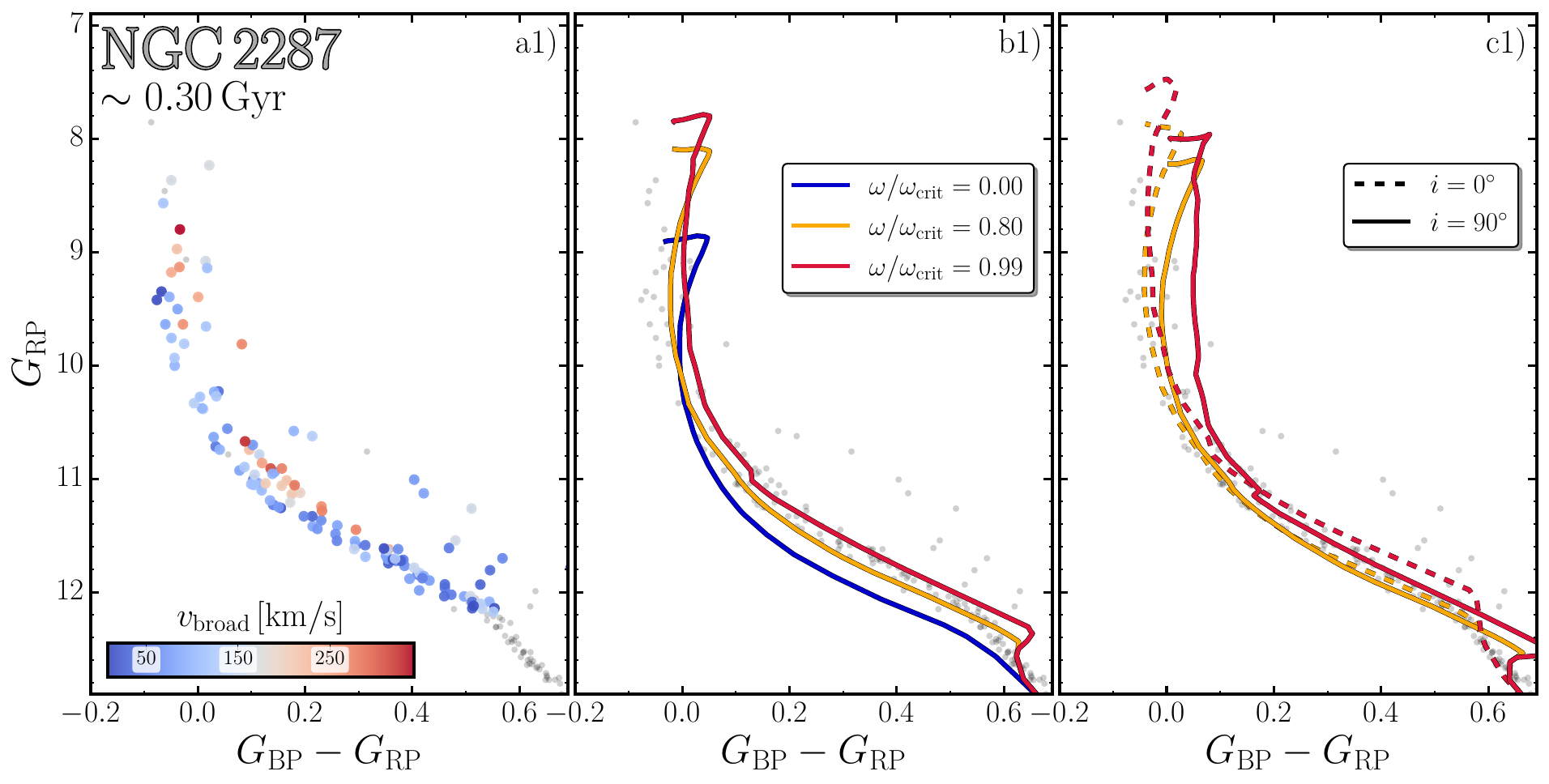}
    \includegraphics[width=0.49\textwidth, trim={0cm 0cm 0cm 0cm}, clip]{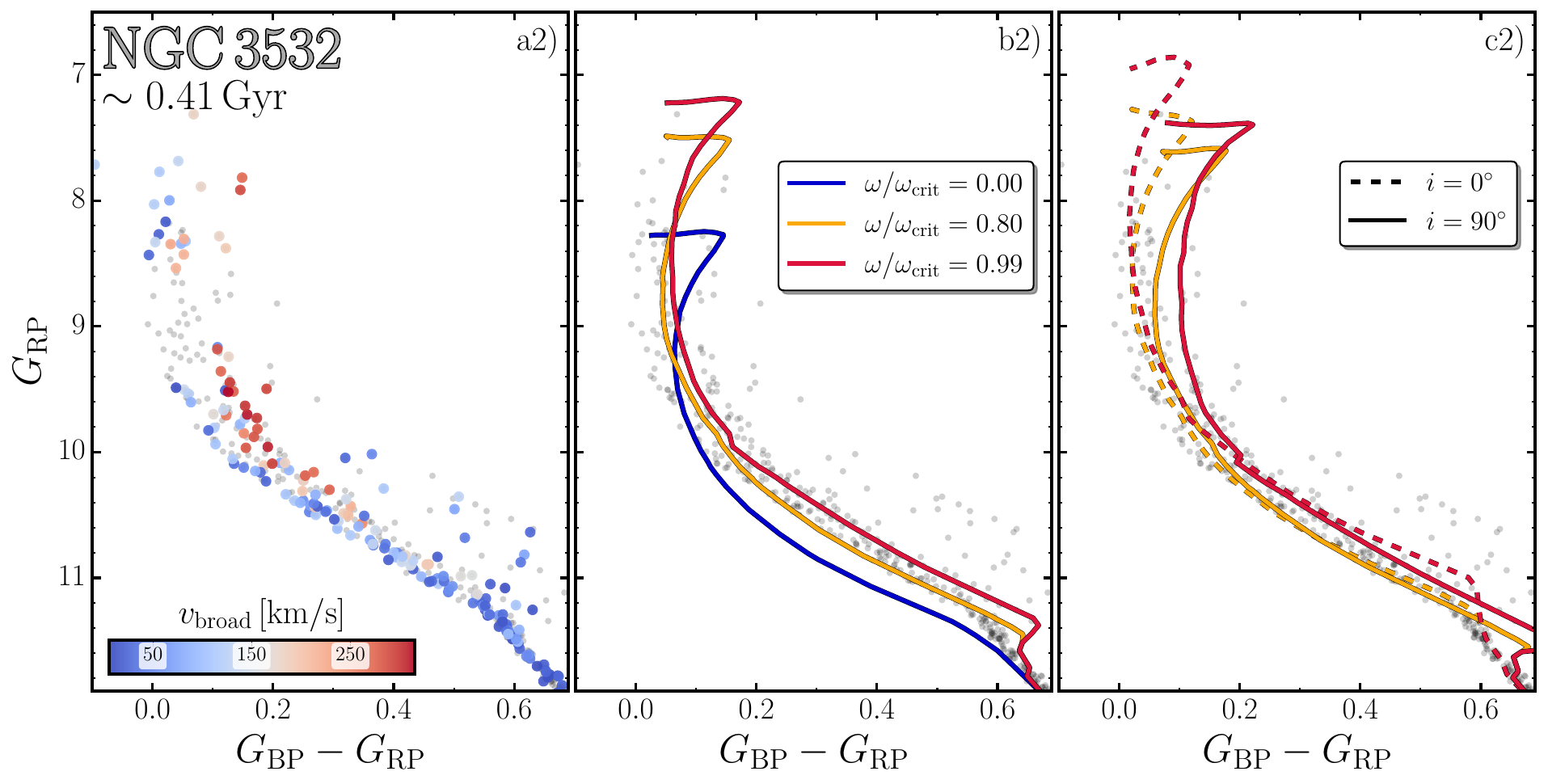}
    \caption{\textit{Panels a).} CMDs of NGC\,2287 (top row) and NGC\,3532 (bottom row), with stars color coded according to the measured Gaia $\vb$, as shown in the bottom colorbar. \textit{Panels b).} Comparison with Parsec theoretical isochrones computed for initial rotation rates as indicated in the right legend, namely 0.00, 0.080 and 0.99 $\omega_\mathrm{crit}$. The displayed isochrones represent a mean computed over all inclinations. \textit{Panels c).} Comparison with Parsec isochrones computed with initial rotation rates as indicated in the legend. Isochrones corresponding to inclinations equal to 0 and 90$^\circ$ are displayed with dashed and solid lines, respectively.}
    \label{fig:splitMS}
\end{figure}

Furthermore, we note that two clusters, namely NGC\,2287 (age$\sim0.3$\,Gyr) and NGC\,3532 (age$\sim0.4$\,Gyr), exhibit hints of bimodality in both color and $\vb$ at the main-sequence level, as shown in Fig.~\ref{fig:splitMS}. Although this pattern is qualitatively similar to observations of young Magellanic Clouds clusters \citep[see e.g.][]{milone2018, dupree2017, marino2018b, kamann2023}, the comparison with Parsec isochrones reveals that the observed split-MSs can be reproduced with initial rotation rates larger than 0.80$\omega_\mathrm{crit}$ and a bimodal spin axis distribution, as can be seen in Fig.~\ref{fig:splitMS}c1-c2. On the other hand, large variations in initial rotation rates would produce a considerable color spread in the MS (see e.g. Fig.~\ref{fig:splitMS}b1-b2), which is not observed. Finally, we remark that NGC\,2287 and NGC\,3532 represent the first two examples of color-$\vsini$ split-MS in young Galactic OCs.


We conclude that, with the current observational dataset and set of theoretical models, we are not able to completely reproduce the properties of young Galactic OCs. Indeed, including a single and fast initial rotation rate can explain the CMDs properties as well as qualitatively reproduce the $\vsini$ vs. $\dc$ relation, but fails to match the observed $\vsini$ distribution. On the other hand, adopting ad-hoc initial rotation distributions may reproduce the $\vsini$ distribution and possibly the $\vsini$ vs. $\dc$ relation, but fails to reproduce the CMDs. Finally, a tailored inclination distribution coupled with a high initial rotation rate might reproduce both the CMD features and the $\vsini$ distribution. \\
However, considering the uncertainties and observational bias (e.g. rotational velocities uncertainties and selection function, and missing binary stars information and possible unknown physical mechanisms), we avoid making any firm conclusions. Nonetheless, we consider the single-rotation interpretation more plausible. 

Nonetheless, at this stage and with the current data, we cannot unequivocally exclude the presence of a small fraction of stars formed non rotating stars.
For example, \citet{dantona2017} showed that stars born as fast rotators and subsequently braked, for instance by magnetic and tidal interactions between binary systems, would exhibit bluer colors and magnitude in the turn-off region. This behavior would hence be consistent with the observed positive correlation between stellar rotation and turn-off color. Therefore, the toy-model discussed here seems to favor a scenario where stars are all born as fast rotators, and/or are possibly subsequently braked, resulting in lower rotational velocities and bluer colors. Additionally, circumstellar dust has been proposed as a possible mechanism contributing to the eMSTO phenomenon in Magellanic Clouds clusters \citep{dantona2023}. Intriguingly, in the context of this work, the presence of dust would contribute to the color spread, possibly loosening the requirements on rotation rates and/or inclination distribution, without affecting our conclusions. Detailed modeling and treatment of interstellar dust in open clusters will be discussed in future works.

\section{Summary and conclusions}
\label{sec:conclusions}
In this work, by combining Gaia DR3 data, i.e. astrometry, photometry and line broadening parameters \citep[$\vb$][]{gaiadr3, fremat2023}, with LAMOST DR9 v2.0 low-resolution spectra \citep{cui2012}, we investigate the extended main-sequence turn-offs in a sample of 32 young Galactic OCs. Specifically, we analyzed the turn-off morphology of all 32 clusters, comparing the observed turn-off morphology with the predictions of rotating models. Moreover, making use Gaia $\vb$ as a proxy of projected stellar rotation, we conduct an homogeneous analysis of stellar rotation among the main-sequence turn-offs in a subset of 17 OCs. Our results, summarized and discussed in the following, provide new and important constraints to understand the origin of the eMSTO phenomenon, confirming the pivotal role of stellar rotation in the evolution of young star clusters. 

First, by comparing projected stellar rotation and chromospheric magnetic activity $(\logrhk)$ for main-sequence and turn-off stars in the sample of analyzed clusters, we confirm the presence of a strong connection between stellar rotation and magnetic activity in F-to-M type stars, with slow rotators exhibiting a lower chromospheric magnetic activity. Conversely, stars with high magnetic activity display on average larger projected rotational velocities. 

Concerning the morphology of the eMSTO and its relation with cluster age, we find that the color spread in the turn-off region increases with cluster age for clusters younger than 2 Gyr. Above this threshold, the observed turn-off color spread significantly decreases to values consistent with photometric uncertainties of Gaia photometry. By comparing our results with Parsec theoretical rotating models, we find that the observed turn-off color spread-age relation is consistent with an initial rotation rates equal to $\omega\sim0.95\omega_\mathrm{crit}$ and distribution of the rotation axis inclination spanning from 0 to 90 degrees. 

Furthermore, exploiting Gaia DR3 $\vb$, we carry out the first direct investigation of the relation between stellar rotation and the eMSTO morphology in 17 open clusters. Our analysis reveals that:
\begin{itemize}
    \item The distribution of projected rotational velocities in the 17 analyzed clusters exhibit hints of bimodality, especially among younger clusters. Such behavior is consistent with previous works \citep{dupree2017,marino2018b, kamann2020, kamann2023}.
    \item Projected rotational velocities $(\vb)$ correlate with the verticalized colors $(\dc)$ of turn-off stars. Specifically, analyzing stars in all clusters simultaneously, we find a Spearman rank coefficients of $R_{\rm S}=0.503 \pm 0.036$. Such results are consistent with the findings discussed in \citet{marino2018a, marino2018b, kamann2020, kamann2023} for Galactic and Magellanic Clouds clusters with similar ages. 
    Moreover, $\vb$-$\dc$ correlation seems to decrease for older clusters, while younger ones exhibit a steeper correlation between $\vb$ and $\dc$.
    \item Comparing the observations with predictions from Parsec theoretical models of rotating stars we find that the properties of CMDs of Open clusters are qualitatively well reproduced by a single population of initially fast rotators (e.g. with initial rotation $\omega \gtrsim 0.90\omega_\mathrm{crit}$) and inclinations encompassing the whole $0^\circ$ and $90^\circ$ range. 
    \item The observed $\vb$ distributions are not well reproduced by a single initial rotation rate and isotropic inclination distribution. Assuming ad-hoc initial rotation rate distributions could reproduce the observed $\vb$, but not the properties of the CMDs, i.e. turn-off and main-sequence spread. On the other and, a single initial rotation rate coupled with bimodal or multi-modal inclination distribution could reproduce the observed CMDs.
    \item Two young clusters, namely NGC\,2287 and NGC\,3532, exhibit hints of split-MSs in both color and $\vb$, with $\vb$ differences as high as $\sim 150$\,km/s. Although such pattern is qualitatively consistent with the observation of young Magellanic Clouds clusters, the small color spread rules out large differences in rotation rate. Additionally, a bimodal spin axis distribution would be required to reproduce the observed split-MSs.
\end{itemize}

In conclusion, our findings suggest that the majority of stars in Galactic Open clusters likely form as fast rotators, with the distribution of rotation axis inclination playing a crucial role in the eMSTO phenomenon. This would represent a major difference from young clusters in the Magellanic Clouds, where a bimodal rotation velocity distribution appears to be mandatory to reproduce the turn-off color spread and split main-sequence. Possible exceptions to this interpretation are represented by the clusters NGC\,2287 and NGC\,3532, where there are some hints of a possible split main-sequence.
On the other hand, at this stage we cannot categorically exclude the presence of a small residual fraction of non rotating stars in Open clusters. Indeed, further investigations will be necessary to fully characterize the nature of stars with low projected rotational velocities in young clusters. Nonetheless, this possible discrepancy between young Open and Magellanic Clouds clusters prompts further investigation into the underlying reasons for the distinct properties of young Galactic and extra-Galactic clusters. Does environment play a role? Or does cluster mass and density contribute to the evolution of the stellar populations in this objects?  

Finally, we advise caution when determining the ages of clusters younger than 1.5 Gyr by means of non-rotating stellar models. Indeed, we find that neglecting the effects stellar rotation in dating open clusters may introduce biases as high as 10-20\%. More efforts are therefore mandatory to quantitatively characterize the effect of including/neglecting stellar rotation in the age derivation of young star clusters, with potential impact on the calibration of the age-rotation-activity relation. 

\section*{Acknowledgments}
This work has received funding from  ``PRIN 2022 2022MMEB9W - \textit{Understanding the formation of globular clusters with their multiple stellar generations}'' (PI Anna F.\,Marino),  and from INAF Research GTO-Grant Normal RSN2-1.05.12.05.10 -  (ref. Anna F. Marino) of the ``Bando INAF per il Finanziamento della Ricerca Fondamentale 2022''

\section*{Data Availability}

Relevant data underlying this article are available in the article and online (\url{https://emsto-ocs.streamlit.app/}). All other data will be shared on reasonable request to the corresponding author.



\bibliographystyle{mnras}
\bibliography{main} 


\newpage

\appendix

\section{Rotating \texttt{Parsec} isochrones} \label{app:parsec}
In the preset work, we have made use of Parsec stellar evolution models which include a detailed treatment of stellar rotation \citep{marigo2017, girardi2019, nguyen2022}. We refer to \citet{girardi2019, nguyen2022} for a detailed description of the physics prescription adopted to compute the isochrones, such as the Roche model for the treatment of stellar geometry and the \citet{espinosa2007} approximation to model the surface $\teff$ distribution. Nonetheless, in this appendix, we provide some details about the properties of the Parsec isochrones which can be useful to better understands the results discussed in the paper.

As briefly mentioned in the text, rotating isochrones are characterized by an initial rotation rate, expressed in units of $\wcrit = (2/3)^{3/2}\sqrt{GM/R^3_\mathrm{pol}}$, where $R_\mathrm{pol}$ indicates the polar radius. Specifically, $\wcrit$ represents the rotation velocity at which the centrifugal force equals the effective gravity at the equator. As discussed in the text, this definition differs from that adopted by MESA models, and can therefore generate discrepancies between different models. For instance, a critical velocity in the Parsec or Geneva models \citep{georgy2013}\footnote{Geneva models use the same definition as Parsec models} corresponds to approximately $\omega = 0.7 \wcrit$ in the MESA models \citep[see e.g][]{wang2023}.
Then, the initial rotation rate of each individual star is evolved up to the age of the requested isochrone. Finally, each isochrones provide stellar magnitudes computed for different inclination of the rotation axis, from 0 to 90 degrees. 

\begin{figure*}
    \centering
    \includegraphics[width=0.75\textwidth, trim={0cm 0cm 0cm 0cm}, clip]{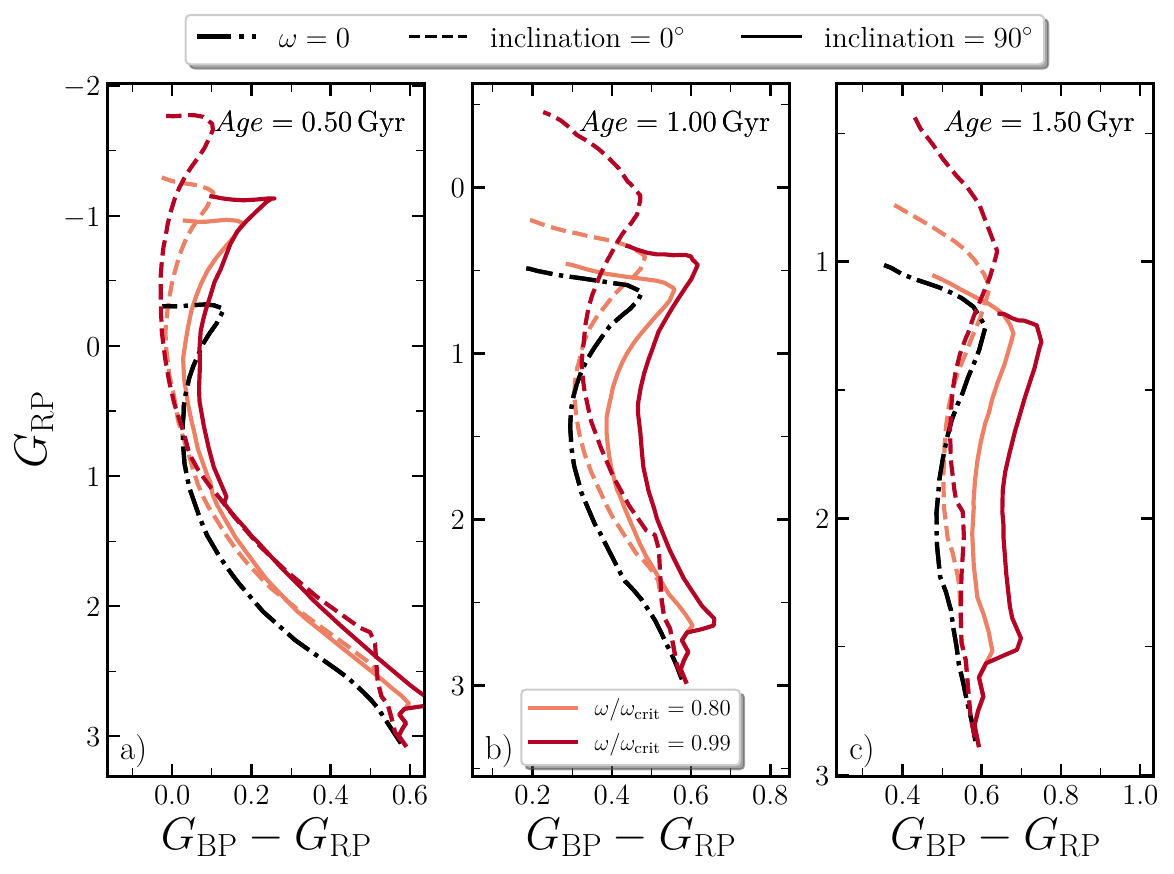}
    \caption{Parsec isochrones with solar metallicity, computed for different ages and initial rotation rates, as indicated in the top and bottom legends. Rotating isochrones with inclination equal to 0 and 90 are respectively shown with colored dashed and solid lines. Comparison non rotating isochrones are shown with black dashed-dotted lines.}
    \label{fig:iso parsec}
\end{figure*}

We show in Fig~.\ref{fig:iso parsec} theoretical isochrones of solar metallicity computed for different initial rotation (orange and red for $\omega=0.80, 0.99\,\wcrit$, respectively) and ages (increasing from left to right). In each panel, the dashed and solid colored lines indicate stars with different initial rotation rate and inclination equal to 0 and 90 degrees, respectively. The dashed-dotted lines represent standard non-rotating isochrone with the same age and metallicity (also shown in blue in the first panel of each row). It can be noted that:
\begin{enumerate}
    \item different inclinations produce a significant color spread at the level of the turn-off, increasing with initial rotation rates and/or ages;
    \item varying the inclination has a small effect on the color of the faint main-sequence, which remain narrow;
    \item standard non rotating isochrones are redder and fainter than rotating ones in clusters younger than 1 Gyr. For older clusters, non rotating isochrones become bluer than rotating ones, exhibiting similar colors and magnitudes to rotating isochrones observed pole-on (i.e. inclination equals to 0 degrees). 
\end{enumerate}

\section{Comparison with literature projected rotational velocities}
\label{app:galah}
To validate the line broadening velocities $(\vb)$ provided by Gaia DR3, similarly to \citet{fremat2023}, we compared Gaia $\vb$ with literature values obtained from GALAH Data Release 3 \cite[DR3][]{buder2021} and \citet[][hereafter referred to as GG]{gg2005} dataset. Specifically, we identified 238 and 416 stars in common between our dataset and the GALAH survey and GG dataset. After applying the Gaia $\vb$ the validation cuts described in \citet{fremat2023}, we are left with 80 and 138 stars in common.

\begin{figure*}
    \centering
    \includegraphics[width=0.99\textwidth, trim={0cm 0cm 0cm 0cm}, clip]{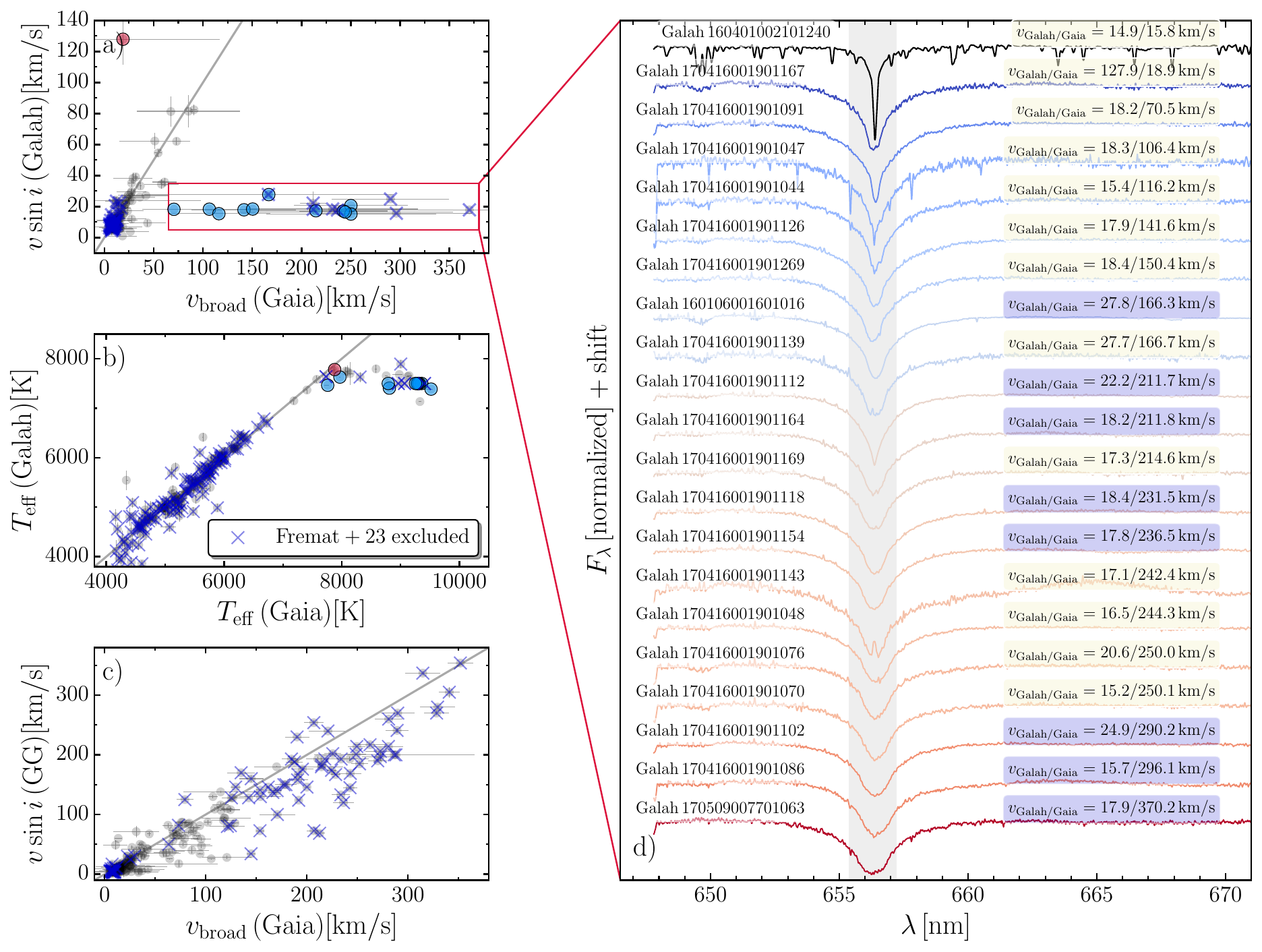}
    \caption{Comparison between Gaia DR3 $\vb$ and literature projected rotation velocities. Blue crosses indicate stars that are excluded according to the criteria discussed in \citet{fremat2023}. \textit{Panel a).} Comparison between Gaia DR3 $\vb$ and GALAH DR3 $\vsini$ \citep{buder2021}. Azure and red circles mark stars with inconsistent values. \textit{Panel b).} Comparison between GALAH DR3 and Gaia DR3 effective temperature. \textit{Panel c).} Comparison between Gaia DR3 $\vb$ and $\vsini$ from \citet{gg2005}. The gray line in each panel indicates the 1:1 relation. \textit{Panel d).} Observed Hermes spectra from GALAH DR3 for the stars enclosed by the red rectangle in panel a., color coded and ordered according to their Gaia $\vb$. Each spectrum has been shifted vertically for a better visualization. GALAH DR3 \texttt{sobject\_id} of each star is listed on the left. The black line in the top corresponds to the spectra  of a star with low rotation in both Gaia and Gala, approximately $15\,\mathrm{km/s}$, and is included for comparison. Spectra of excluded stars are marked by blue shaded right insets.}
    \label{fig:comp galah}
\end{figure*}

The comparison, illustrated in Fig.\ref{fig:comp galah}, unveils a good agreement between Gaia DR3 $\vb$ and the literature values. Indeed, the majority of Gaia $\vb$ values align closely with those reported in \citet{gg2005}, with a minority exhibiting slightly larger values. The cross-match with GALAH DR3 reveals that 86\% of stars exhibit consistent values, while the remaining 16\% of stars (azure circles in Fig.~\ref{fig:comp galah}) have larger Gaia $\vb$ compared to their GALAH counterparts. Interestingly, one star, highlighted as a red circle, displays the opposite trend, with GALAH velocity exceeding the Gaia measurement.

However, we note that stars with inconsistent Gaia/GALAH velocities display very high temperatures, e.g., $T_\mathrm{eff}>8000,\mathrm{K}$, as determined by Gaia. Such temperatures approach the upper limits of GALAH's temperature range\citep[see e.g. Fig. 3 of][]{buder2021}, potentially explaining the observed discrepancies. To support this assertion, we provide Hermes red arm spectra\footnote{We display the pseudo-continuum normalized spectra with sky subtracted, sourced from GALAH DR3 \citep[see][for further details]{buder2021}} of stars with inconsistent velocities (azure and red circles in panels a-b), where the presence of the $H_\mathrm{\alpha}$ line is evident. Notably, while second-order effects might contribute to line broadening, stellar rotation is the primary source. Hence, the broadening of the $H_\mathrm{\alpha}$ line serves as a powerful diagnostic for determining projected stellar rotation \citep[see e.g.][]{dupree2017, marino2018a, marino2018b}. For comparison, we include the spectra of a star with low rotation velocity, approximately $\sim 15,\mathrm{km/s}$ in all datasets, depicted as a black line in the top of Fig.~\ref{fig:comp galah}d. GALAH \texttt{sobject\_id}s are indicated on the left side of each spectrum. Remarkably, a visual inspection of the spectra reveals that the $H_\mathrm{\alpha}$ line in stars with inconsistent $\vb$ values consistently appears broader than that of the comparison slow-rotator (black line), strongly indicating the fast-rotating nature of these stars, hence confirming Gaia DR3 $\vb$ measurements. On the other hand, the star with larger GALAH velocity, i.e. red circle in panels a-b and second spectra from the top, is consistent with being a fast-rotator, as determined by GALAH.

Consequently, the substantial agreement between Gaia DR3 $\vb$ and literature values from GALAH DR3 (or Hermes spectra) and \citet{gg2005}, confirms the reliability and accuracy of Gaia $\vb$ as proxies for projected stellar rotation, i.e., $\vsini$.

Additionally, we find worth noting that at least one star in the sample whose spectra is plotted in Fig.~\ref{fig:comp galah}d exhibit sign of emisssion in the core of the $H_\mathrm{\alpha}$ line, namely Gaia DR3 3064634473378653440/GALAH DR3 170416001901143. The temperature of this star, as determined by Gaia, is $\teff \sim 9000$\,K. Hence we speculate it could represent an emitting A-star, or Ae-star.

\bsp	
\label{lastpage}
\end{document}